\documentclass[journal]{IEEEtran}

\usepackage{xcolor}

\usepackage{cite}

\usepackage{amsmath,amssymb,amsfonts}

\usepackage{algorithmic}

\usepackage{array}

\usepackage{subcaption}
\usepackage[pdftex]{graphicx}

\usepackage{url}

\usepackage[us]{datetime} 
\usepackage{supertabular}
\usepackage{eurosym}
\newtheorem{demonstration}{Demonstration}
\usepackage[draft=False]{hyperref} 
\hypersetup{
     colorlinks=true,
     linkcolor=black,
     filecolor=black,
     citecolor=black,      
     urlcolor=black,
     }

\newcommand{\forallt}{\ensuremath{\forall t \in \mathcal{T}}}

\newcommand{\convdate}{\formatdate{14}{09}{2019}}

\begin{document}
%

\title{A Probabilistic Forecast-Driven Strategy for a Risk-Aware Participation in the Capacity Firming Market: extended version}
%
%
%

\author{Jonathan~Dumas,
        Colin~Cointe,
        Antoine~Wehenkel,
        Antonio~Sutera,
        Xavier~Fettweis,
        and~Bertrand~Corn\'elusse
\thanks{The authors are with the Departments of Computer Science and Electrical Engineering and Geography, University of Li\`ege, 4000 Li\`ege, Belgium, (e-mail: \{jdumas, antoine.wehenkel, a.sutera, xavier.fettweis, bertrand.cornelusse\}@uliege.be, colin.cointe@mines-paristech.fr).}
}

%
%

\markboth{IEEE Transactions on Sustainable Energy}%
{Shell \MakeLowercase{\textit{et al.}}: Bare Demo of IEEEtran.cls for IEEE Journals}
%

\maketitle


\begin{abstract}
This paper addresses the energy management of a grid-connected renewable generation plant coupled with a battery energy storage device in the capacity firming market, designed to promote renewable power generation facilities in small non-interconnected grids.
The core contribution is to propose a probabilistic forecast-driven strategy, modeled as a min-max-min robust optimization problem with recourse. It is solved using a \textit{Benders-dual cutting plane} algorithm and a \textit{column and constraints generation} algorithm in a tractable manner. A dynamic risk-averse parameters selection strategy based on the quantile forecasts distribution is proposed to improve the results. 
A secondary contribution is to use a recently developed deep learning model known as \textit{normalizing flows} to generate quantile forecasts of renewable generation for the robust optimization problem. This technique provides a general mechanism for defining expressive probability distributions, only requiring the specification of a base distribution and a series of bijective transformations.
Overall, the robust approach improves the results over a deterministic approach with nominal point forecasts by finding a trade-off between conservative and risk-seeking policies.
The case study uses the photovoltaic generation monitored on-site at the University of Li\`ege (ULi\`ege), Belgium. 
\end{abstract}


\section{Notation}\label{sec:notation}
\subsection*{Sets and indices}
\begin{supertabular}{l p{0.7\columnwidth}}
	Name & Description \\
	\hline
	$t$ & Time period index. \\
	$T$ & Number of time periods per day. \\
	$\mathcal{T}$ & Set of time periods, $\mathcal{T}= \{1,2, \ldots, T\}$. \\
	$\mathcal{P}$  &  Renewable generation uncertainty set. \\
\end{supertabular}
\subsection*{Variables}
\begin{supertabular}{l l p{0.45\columnwidth}}
	Name & Range & Description \\
	\hline
	$x_t$ & $ [X_t^{min},X_t^{max}]$ & Engagement, kW. \\
	$y_t$ & $ [Y_t^{min},Y_t^{max}]$ & Net power, kW. \\
	$y_t^G $ & $ [0, P_c]$ & Renewable generation, kW.  \\
	$y_t^{cha} $ & $ [0,S^c]$ & Charging power, kW.    \\
	$y_t^{dis} $ & $ [0,S^d]$ & Discharging power, kW.   \\
	$y_t^s$ & $ [S^{min}, S^{max}]$ & BESS state of charge, kWh. \\
	$\delta x_t^-$, $\delta x_t^+$  & $ \mathbb{R}_+$ & Under/overproduction, kW.  \\
	$y_t^b$ & $ \{0, 1\}$ & BESS binary variable. \\
	$z_{t}$ & $ \{0, 1\}$ & Uncertainty set binary variable. \\
	$\alpha_t$ & $ [M_t^-,M_t^+]$ & Variables to linearize $z_t \phi^{y^G}_t$.   \\
\end{supertabular}
\subsection*{Dual variables, and corresponding constraints}
\noindent Dual variables of constraints are indicated with brackets $[ \cdot ]$.
\begin{supertabular}{l l p{0.5\columnwidth} }
	Name & Range & Description \\
	\hline
	$\phi^{cha}_t $, $\phi^{dis}_t $ & $ \mathbb{R}^-$ & Maximum storage (dis)charging.  \\
	$\phi^{S^{min}}_t $, $\phi^{S^{max}}_t $ & $ \mathbb{R}^-$ & Minimum/maximum storage capacity.   \\
	$\phi^y_t$ & $ \mathbb{R}$ & Net power balance. \\
	$\phi^{Y^{min}}_t$, $\phi^{Y^{max}}_t$ & $ \mathbb{R}^-$ & Minimum/maximum net power. \\
	$\phi^{S^i}_t $, $\phi^{S^f}_t $ & $ \mathbb{R}^-$ & Initial/final state of charge.   \\
	$\phi^{y^s}_t$ & $ \mathbb{R}^-$ & BESS dynamics. \\
	$\phi^{\delta x^-}_t$, $\phi^{\delta x^+}_t$ & $ \mathbb{R}^-$ & Under/overproduction.  \\
	$\phi^{y^G}_t$ & $ \mathbb{R}^-$ & Renewable generation.  \\
\end{supertabular}
\subsection*{Parameters}
\begin{supertabular}{l p{0.65\columnwidth} }
	Name & Description  \\
	\hline
	$X_t^{min}$, $X_t^{max}$ & Minimum/maximum engagement, kW. \\
	$\Delta X_t$  & Ramping-up and down limits for the engagement, kW.  \\
	$p P_c$ & Engagement tolerance, $ 0 \leq p \leq 1$, kW.  \\
	$y_t^m$ & Net measured power, kW. \\
	$Y_t^{min}$, $Y_t^{max}$ & Minimum/maximum net power, kW. \\
	$P_c$ & Total installed capacity, kWp.  \\
	$\hat{p}_t$, $\hat{p}^{(q)}_t$ & Point/quantile $q$ forecast, kW.  \\
	$p^{min}_t$, $p^{max}_t$ & Uncertainty set lower/upper bounds, kW.  \\
	$S^d$, $S^c$ & BESS maximum (dis)charging power, kW.  \\
	$\eta^d$, $\eta^c$ & BESS (dis)charging efficiency. \\
	$S^{min}$, $S^{max}$ & BESS minimum/maximum capacity, kWh. \\
	$S^i$, $S^f$ & BESS initial/final state of charge, kWh.  \\
	$\pi_t$ & Contracted selling price, \euro/kWh.   \\
	$\Delta t$ & Duration of a time period, minutes.   \\
	$\Gamma$ & Uncertainty budget.  \\
	$\beta$ & Penalty factor.  \\
	$d_q$, $d_\Gamma$ & Uncertainty and budget depths.  \\
	$M_t^-$, $M_t^+$ & Big-M’s values.  \\
\end{supertabular}

\section{Introduction}

\IEEEPARstart{T}{he} capacity firming framework is mainly designed for isolated markets, such as the Overseas France islands. For instance, the French Energy Regulatory Commission (CRE) publishes capacity firming tenders and specifications. The system considered is a grid-connected renewable energy power plant, \textit{e.g.}, photovoltaic or wind-based, with a battery energy storage system (BESS) for firming the renewable generation. At the tendering stage, offers are selected on the electricity selling price. Then, the successful tenderer builds its plant and sells the electricity exported to the grid at the contracted selling price, but according to a well-defined daily engagement and penalization scheme specified in the tender specifications. The electricity injected in or withdrawn from the grid must be nominated the day-ahead, and engagements must satisfy ramping power constraints. The remuneration is calculated a posteriori by multiplying the realized exports by the contracted selling price minus a penalty. The deviations of the realized exports from the engagements are penalized through a function specified in the tender. A peak option can be activated in the contract for a significant selling price increase during a short period defined a priori. Therefore, the BESS must shift the renewable generation during peak hours to maximize revenue and manage renewable energy uncertainty.

The problem of modeling a two-phase engagement/control with an approach dealing with uncertainty in the context of the CRE capacity framework is still an open issue. This framework has received less attention in the literature than more traditional energy markets such as day-ahead and intraday markets of European countries.
There are several approaches to deal with renewable energy uncertainty. One way is to consider a two-stage stochastic programming approach\cite{birge2011introduction}, that has already been applied to the capacity firming framework \cite{dumas2020probabilistic,n2020controle,haessig2014dimensionnement,parisio2016stochastic}. 
The generation uncertainty is captured by a set of scenarios modeling possible realizations of the power output. However, this approach has three drawbacks. First, the problem size and computational requirement increase with the number of scenarios, and a large number of scenarios are often required to ensure the good quality of the solution. Second, the accuracy of the algorithm is sensitive to the scenario generation technique. Finally, it may be challenging to identify an accurate probability distribution of the uncertainty.
Another option is to consider robust optimization (RO)\cite{ben2009robust,bertsimas2011theory}, applied to unit commitment by \cite{bertsimas2012adaptive,jiang2011robust}, and in the capacity firming setting \cite{n2020controle}. RO accounts for the worst generation trajectory to hedge the power output uncertainty, where the uncertainty model is deterministic and set-based. Indeed, the RO approach puts the random problem parameters in a predetermined uncertainty set containing the worst-case scenario.
It has two main advantages \cite{bertsimas2012adaptive}: (1) it only requires moderate information about the underlying uncertainty, such as the mean and the range of the uncertain data; (2) it constructs an optimal solution that immunizes against all realizations of the uncertain data within a deterministic uncertainty set. Therefore, RO is consistent with the risk-averse fashion way to operate power systems.
However, the RO version of a tractable optimization problem may not itself be tractable, and some care must be taken in choosing the uncertainty set to ensure that tractability is preserved.

Traditionally, a two-stage RO model is implemented for the unit commitment problem in the presence of uncertainty. However, it is challenging to compute and often NP-hard. 
Two classes of cutting plane strategies have been developed to overcome the computational burden. The Benders-dual cutting plane (BD) algorithms are the most used and seek to derive exact solutions in the line of Benders’ decomposition \cite{benders1962partitioning} method. They decompose the overall problem into a master problem involving the first-stage commitment decisions at the outer level and a sub-problem associated with the second-stage dispatch actions at the inner level. Then, they gradually construct the value function of the first-stage decisions using dual solutions of the second-stage decision problems \cite{bertsimas2012adaptive,jiang2011robust}.
In contrast, the column-and-constraint generation (CCG) procedure, introduced by \cite{zhao2012robust,zeng2013solving} does not create constraints using dual solutions of the second-stage decision problem. Instead, it dynamically generates constraints with recourse decision variables in the primal space for an identified scenario. The generated variables and constraints in the CCG procedure are similar to those in the deterministic equivalent of a two-stage stochastic programming model.
The BD and CCG algorithms have not been compared in the capacity firming framework to the best of our knowledge.

\begin{figure}[tb]
	\centering
	\includegraphics[width=\linewidth]{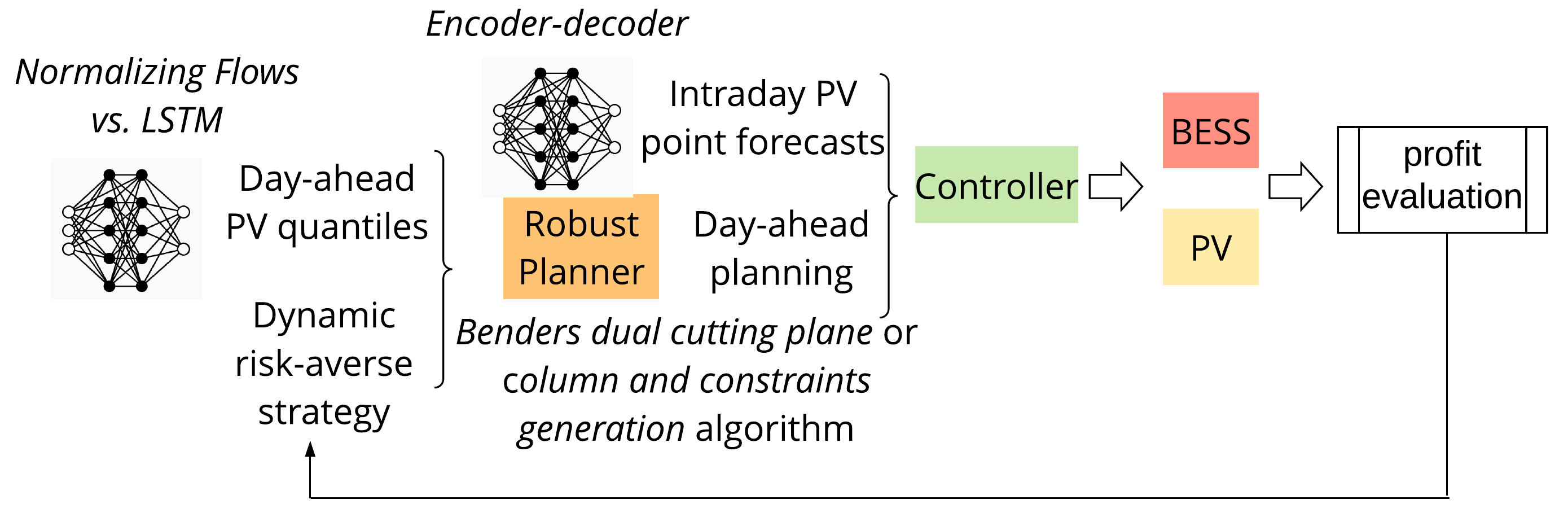}
	\caption{Forecast-driven robust optimization strategy.}
	\label{fig:process}
\end{figure}

This paper proposes a reliable and computationally tractable probabilistic forecast-driven robust optimization strategy. It can use either a BD or CGG algorithm in the capacity firming framework, depicted in Figure \ref{fig:process}. 
Our work goes several steps further than \cite{n2020controle}. The main contributions of this paper are three-fold:
\begin{enumerate}
\item The core contribution is applying the robust optimization framework to the capacity firming market in a tractable manner using a Benders decomposition. The non-linear robust optimization problem is solved both using the Benders-dual cutting plane and the column-and-constraint generation algorithms. To the best of our knowledge, it is the first time these algorithms are compared in the capacity firming framework.
In addition, the convergence of the BD algorithm is improved with a warm-start procedure. It consists of building an initial set of cuts based on renewable generation trajectories assumed to be close to the worst-case scenario.
The results of both the CCG and BD two-stage RO planners are compared to the deterministic planner using perfect knowledge of the future, the nominal point forecasts, \textit{i.e.}, the baseline to outperform, and the quantiles (a conservative approach). The case study is the photovoltaic (PV) generation monitored on-site at the University of Li\`ege (ULi\`ege), Belgium.
%
\item Second, a dynamic risk-averse parameters selection taking advantage of the quantile forecast distribution is investigated and compared to a strategy with fixed risk-averse parameters.
%
\item Finally, the \textit{normalizing flows} (NFs) is implemented. In recent years, a new class of probabilistic generative models has gained increasing interest from the deep learning community. NFs are used to compute day-ahead quantiles of renewable generation for the robust planner. Then, an encoder-decoder architecture forecasting model \cite{bottieau2019very} computes the intraday point forecasts for the controller. To the best of our knowledge, it is the first study to use NFs in a power system application.
\end{enumerate}

In addition to these contributions, this study also provides open-access to the Python code\footnote{\url{https://github.com/jonathandumas/capacity-firming-ro}} to help the community to reproduce the experiments.
The rest of the paper is organized as follows. Section~\ref{sec:capacity_firming_process} describes the capacity firming framework. Section~\ref{sec:formulation} provides the mathematical formulations of the robust and deterministic planners. Section~\ref{sec:ro_benders} develops the Benders decomposition algorithm used to solve the robust formulation. The case study and computational results are shown in Section~\ref{sec:case_study}. Section~\ref{sec:conclusions} concludes our research and draws some perspectives of future works. 
Appendix~\ref{appendix:forecasting} introduces the forecasting techniques and proposes a quality evaluation. 
Appendix~\ref{appendix:warm-start} presents the BD warm-start procedure. Finally, Appendix~\ref{appendix:ccg} details the column and constraints generation algorithm implemented.

\section{The Capacity Firming Framework}\label{sec:capacity_firming_process}

The capacity firming framework can be decomposed into a day-ahead engagement process, Section \ref{sec:engagement_process}, and a real-time control process, Section \ref{sec:control_process}. Each day is discretized in $T$ periods of duration $\Delta t$. In the sequel, the period duration is the same for day-ahead engagement and the real-time control, $t$ is used as a period index, and $\mathcal{T}$ is the set of periods in a day. 

\subsection{Day-ahead engagement}\label{sec:engagement_process}

Each day, the renewable generation plant operator has to provide the day-ahead production profile to the grid operator, based on renewable generation forecasts. More formally, a planner computes on a day-ahead basis, before a deadline, a vector of engagements composed of $T$ values $\{ x_1,...,x_T \}$. The grid operator accepts the engagements if they satisfy the constraints
\begin{subequations}\label{eq:engagement_constraints}
	\begin{align}
	|x_t-x_{t-1}| & \leq  \Delta X_t , \ \forallt \setminus \{1\}	\\
	- x_t &  \leq -  X_t^{min}, \ \forallt  \\
	x_t &  \leq X_t^{max}, \ \forallt ,
	\end{align}
\end{subequations}
where $\forallt \setminus \{1\}$ is $\mathcal{T}$ without the first period, $\Delta X_t$ a power ramping constraint, that is a fraction of the total installed capacity $P_c$ determined at the tendering stage and imposed by the grid operator. 

\subsection{Real-time control}\label{sec:control_process}

Then, in real-time, a receding-horizon controller computes at each period the generation level and the charge or discharge set-points from $t$ to $T$, based on forecasts of renewable generation and the engagements. Only the set-points of the first period are applied to the system.
The remuneration is calculated ex-post based on the realized power $y^m_t$ at the grid coupling point. For a given control period, the net remuneration $r_t$ of the plant is the gross revenue $ \Delta t \pi_t  y^m_t$ minus a penalty $c(x_t, y^m_t)$, with $\pi_t$ the contracted selling price set at the tendering stage
\begin{align}\label{eq:remuneration}	
r_t = \Delta t \pi_t  y^m_t  - c(x_t, y^m_t),  \ \forallt.
\end{align}
The penalty function $c$ depends on the specifications of the tender. For the sake of simplicity in the rest of the paper, $c$ is assumed to be symmetric, convex, and piecewise-linear.

\section{Optimization problems formulation}\label{sec:formulation}

A two-stage robust optimization formulation is adopted to deal with the engagement for the uncertain renewable generation. The deterministic and robust formulations of the planner are presented in Sections \ref{sec:det_planner} and \ref{sec:ro_planner}. The robust optimization problem with recourse has the general form of a min-max-min optimization problem. An uncertainty set defined by quantiles forecasts and a budget of uncertainty $\Gamma$ models the renewable generation. Section \ref{sec:ro_planner_transformed} uses the dual of the inner problem to formulate a min-max optimization problem. Finally, Section \ref{sec:controller} presents the formulation of the controller.

\subsection{Two-stage deterministic planner formulation}\label{sec:det_planner}

The objective function $J $ to minimize is the opposite of the net revenue
\begin{align}\label{eq:objective}
J \big(x_t, y_t \big) & \ =  \sum_{t\in \mathcal{T}}\pi_t \Delta t  [-y_t + \beta (\delta x_t^- + \delta x_t^+)] ,
\end{align}
with $\beta$ a penalty factor. The deterministic formulation is the following Mixed-Integer Linear Program (MILP)
\begin{align}\label{eq:det_compact}
\min_{x_t \in \mathcal{X}, y_t \in \Omega(x_t, \hat{p}_t)}& \ J \big(x_t, y_t \big)  ,
\end{align}
where $$\mathcal{X} =  \big \{ x_t : (\ref{eq:MILP_engagement_cst}) \big \}$$ and $$\Omega(x_t, \hat{p}_t) = \big \{ y_t : (\ref{eq:MILP_BESS_charge_discharge_cst})- (\ref{eq:MILP_G})  \big \}$$ are the sets of feasible engagements $x_t$ and dispatch solutions $y_t$ for a fixed engagement $x_t$ and renewable generation point forecast $\hat{p}_t$. The optimization variables of (\ref{eq:det_compact}) are the engagement variables $x_t$, the dispatch variables $y_t$ (the net power at the grid connection point), $y_t^{dis}$ (discharging power), $y_t^{cha}$ (charging power), $y_t^s$ (BESS state of charge), $y_t^{b}$ (BESS binary variables), $y_t^G$ (renewable generation), and $\delta x_t^-, \delta x_t^+$ (threshold-linear penalty variables) (cf. Section~\ref{sec:notation}).
From (\ref{eq:engagement_constraints}), the engagement constraints are\footnote{The ramping constraint on $x_1$ is deactivated to decouple consecutive days of simulation. In reality, the updated value of the last engagement of the previous day would be taken to satisfy the constraint.}
\begin{subequations}
	\label{eq:MILP_engagement_cst}	
	\begin{align}
	x_t-x_{t-1} & \leq  \Delta X_t, \ \forallt \setminus \{1\}  \\
	x_{t-1}-x_t & \leq   \Delta X_t, \ \forallt \setminus \{1\}\\
	- x_t &  \leq -  X_t^{min}, \ \forallt  \\
	x_t &  \leq X_t^{max}, \ \forallt.
	\end{align}
\end{subequations}
The set of constraints that bound $y_t^{cha}$, $y_t^{dis}$, and $y_t^s$ variables are $\forallt$ 
\begin{subequations}\label{eq:MILP_BESS_charge_discharge_cst}	
	\begin{align}
	y_t^{cha}  & \leq  y_t^{b} S^c & [\phi^{cha}_t] \label{eq:MILP_BESS_charge_discharge_cst1}	 \\
	y_t^{dis}  & \leq (1- y_t^{b}) S^d  & [\phi^{dis}_t] \label{eq:MILP_BESS_charge_discharge_cst2} 	\\
	- y_t^s & \leq -S^{min} & [\phi^{S^{min}}_t] \label{eq:MILP_BESS_charge_discharge_cst3}	 \\
	y_t^s & \leq  S^{max}, & [\phi^{S^{max}}_t] \label{eq:MILP_BESS_charge_discharge_cst4}	
	\end{align}
\end{subequations}
where $y_t^{b}$ are binary variables that prevent the simultaneous charge and discharge of the BESS. The power balance equation and the constraints on the net power at the grid connection point are $\forallt$ 
\begin{subequations}\label{eq:MILP_balance}	
	\begin{align}
	y_t - y_t^G - \left(y_t^{dis} - y_t^{cha} \right) &  =   0 & [\phi^y_t] \label{eq:MILP_balance_1}	\\
	- y_t&  \leq - Y_t^{min}   \label{eq:MILP_balance_2}  & [\phi^{Y^{min}}_t] \\
	y_t &  \leq Y_t^{max} . \label{eq:MILP_balance_3}	 & [\phi^{Y^{max}}_t] 
	\end{align}
\end{subequations}
The dynamics of the BESS state of charge are \footnote{The parameters $S^f$ and $S^i$ are introduced to decouple consecutive days of simulation. In reality, $S^i$ would be the updated value of the last measured state of charge of the previous day.}
\begin{subequations}\label{eq:MILP_soc_dynamic}
	\begin{align}
	&y^s_1 - \Delta t (  \eta^c  y^{cha}_1 - \frac{y^{dis}_1}{\eta^d}   ) = S^i & [\phi^{S^i}] \label{eq:MILP_soc_dynamic1}\\
	&y^s_t - y^s_{t-1} - \Delta t (  \eta^c  y^{cha}_t - \frac{y^{dis}_t}{\eta^d}  )=0 , \notag \\ 
	& \quad \forallt \setminus \{1\}  & [\phi^{y^s}_t] \label{eq:MILP_soc_dynamic2}\\
	&y^s_{T} = S^f = S^i.  & [\phi^{S^f}] \label{eq:MILP_soc_dynamic3}
	\end{align}
\end{subequations}
The variables $\delta x_t^-, \delta x_t^+ $ are defined $\forallt$ to model the symmetric threshold-linear penalty
\begin{subequations} \label{eq:MILP_penalty_csts}	
	\begin{align}
	- \delta x_t^- & \leq     \big( y_t -(x_t - p P_c) \big) & [\phi^{\delta x^-}_t] \label{eq:MILP_penalty_1}\\
	- \delta x_t^+, & \leq     \big( (x_t + p P_c) - y_t \big), & [\phi^{\delta x^+}_t]\label{eq:MILP_penalty_2}
	\end{align}
\end{subequations}
with $ 0\leq p \leq 1$.
Finally, the renewable generation is bounded by the point forecast $\hat{p}_t$ $\forallt$
\begin{align}\label{eq:MILP_G}	
y_t^G & \leq  \hat{p}_t . & [\phi^{y^G}_t]
\end{align}

\subsection{Two-stage robust planner formulation}\label{sec:ro_planner}

The uncertain renewable generation $\hat{p}_t$ of (\ref{eq:MILP_G}) is assumed to be within an interval $[p_t^{min}, p_t^{max}]$ that can be obtained based on the historical data or an interval forecast composed of quantiles.
In the capacity firming framework, where curtailment is allowed,  the uncertainty interval consists only in downward deviations $[p_t^{min}, \hat{p}^{(0.5)}_t]$, with $\hat{p}^{(0.5)}_t$ the 50 \% quantile.
\begin{demonstration}\label{demonstration}
Consider $\mathcal{P}_1 =  [ p_t^{min}, \hat{p}^{(0.5)}_t]$, $\mathcal{P}_2 = [ \hat{p}^{(0.5)}_t, p_t^{max}]$, $\hat{p}_t^1 \in \mathcal{P}_1$, and $ \hat{p}_t^2 \in \mathcal{P}_2$. It is obvious that $\hat{p}_t^1 \leq \hat{p}_t^2$ $\forallt$. Then, consider $y_t^1 \in  \Omega_1 =  \Omega(x_t, \hat{p}_t^1 \in \mathcal{P}_1) $ and $y_t^2 \in  \Omega_2 =  \Omega(x_t, \hat{p}_t^2 \in \mathcal{P}_2) $. The only difference between $\Omega_1$ and $\Omega_2$ is (\ref{eq:MILP_G}) where $y_t^{G,1} \leq \hat{p}_t^1$ and $y_t^{G,2} \leq \hat{p}_t^2$. However, as $\hat{p}_t^1 \leq \hat{p}_t^2$, it is straightforward that $y_t^{G,1} \leq \hat{p}_t^2$. Therefore, $y_t^1 \in \Omega_2$ $\forallt$, and $\Omega_1 \subseteq \Omega_2$. Thus, $\min_{y_t \in \Omega_2} J (x_t, y_t ) =J_2^\star \leq \min_{y_t \in \Omega_1} J (x_t, y_t ) =J_1^\star$, and $\max(J_2^\star, J_1^\star) = J_1^\star$. It means the worst case is in $\Omega_1$ that corresponds to $\mathcal{P}_1$.
\end{demonstration}

In addition, the worst generation trajectories, in robust unit commitment problems, are achieved when the uncertain renewable generation $\hat{p}_t$ reaches the lower or upper bounds of the uncertainty set (Proposition 2 of \cite{zhao2012robust}). Thus, the uncertainty set at $t$ is composed of two values and $\hat{p}_t \in \{p_t^{min}; \hat{p}^{(0.5)}_t \}$.
Following \cite{bertsimas2012adaptive,jiang2011robust}, to adjust the degree of conservatism, a budget of uncertainty $\Gamma$ taking integer values between 0 and 95 is employed to restrict the number of periods that allow $\hat{p}_t$ to be far away from its nominal value, \textit{i.e.}, deviations are very large.
Therefore, the uncertainty set of renewable generation $\mathcal{P}$ is defined as follows
\begin{align}\label{eq:G_set}	
\mathcal{P} =   \big \{p_t \in \mathbb{R}^{T} : & \sum_{t\in \mathcal{T}}  z_t  \leq \Gamma , \ z_t \in \big \{0;1 \big \} \ \forallt,\notag \\
& p_t  = \hat{p}^{(0.5)}_t - z_t p_t^{min} \ \forallt \big \} ,
\end{align}
where $p_t^{min} = \hat{p}^{(0.5)}_t - \hat{p}^{(q)}_t$, with $0 \leq q \leq 0.5$. When $\Gamma = 0$, the uncertainty set $\mathcal{P} =\{ \hat{p}^{(0.5)}_t \} $ is a singleton, corresponding to the nominal deterministic case. As $\Gamma$ increases the size of $\mathcal{P}$ enlarges. This means that a larger total deviation from the expected renewable generation is considered, so that the resulting robust solutions are more conservative and the system is protected against a higher degree of uncertainty. When $\Gamma = T$, $\mathcal{P}$ spans the entire hypercube defined by the intervals for each $p_t$.
%
With this uncertainty set description, the proposed two-stage robust formulation of the capacity firming problem consists of minimizing the objective function over the worst renewable generation trajectory
\begin{align}\label{eq:RO_min_max}	
\max_{\hat{p}_t \in \mathcal{P}} \bigg[\min_{x_t \in \mathcal{X}, \ y_t \in \Omega(x_t, \hat{p}_t)}  &  \  J \big(x_t, y_t \big) \bigg],
\end{align}
that is equivalent to
\begin{align}\label{eq:RO_min_max_min}	
\min_{x_t \in \mathcal{X}} \bigg[ \max_{\hat{p}_t \in \mathcal{P}} \min_{y_t \in \Omega(x_t, \hat{p}_t)} & \ J \big(x_t, y_t \big) \bigg] .
\end{align}
The worst-case dispatch cost has a max-min form, where $$\min_{y_t \in \Omega(x_t, \hat{p}_t)}  J \big(x_t, y_t \big)$$ determines the economic dispatch cost for a fixed engagement and a renewable generation trajectory, which is then maximized over the uncertainty set $\mathcal{P}$.

\subsection{Second-stage planner transformation}\label{sec:ro_planner_transformed}

The proposed formulation (\ref{eq:RO_min_max_min}) consists of solving a min-max-min problem, which cannot be solved directly by a commercial software such as CPLEX or GUROBI. A scenario-based approach, \textit{e.g.}, enumerating all possible outcomes of $\hat{p}_t$ that could lead to the worst-case scenario for the problem, results in at least $2^\Gamma$ possible trajectories\footnote{There are $n =\sum_{k=0}^{\Gamma}  \binom{96}{k} $ possible trajectories where $n$ is within the interval $[2^\Gamma, 2^{96}]$ as $(1+1)^\Gamma = \sum_{k=0}^{\Gamma}  \binom{\Gamma}{k}$ and $(1+1)^{96} = \sum_{k=0}^{96}  \binom{96}{k}$ by using the binomial formula.}. Thus, to deal with the huge size of the problem a Benders type decomposition algorithm is implemented.

Constraints (\ref{eq:MILP_BESS_charge_discharge_cst1})-(\ref{eq:MILP_BESS_charge_discharge_cst2}) make the dispatch problem a MILP, for which a dual formulation cannot be derived. In view of this, following \cite{jiang2011robust}, the constraints (\ref{eq:MILP_BESS_charge_discharge_cst1})-(\ref{eq:MILP_BESS_charge_discharge_cst2}) are relaxed (the convergence of the relaxed dispatch problem is discussed in Section \ref{sec:convergence-checking}).
Then, by applying standard tools of duality theory in linear programming, the constraints and the objective function of the dual of the dispatch problem are derived.
The dual of the feasible set $\Omega(x_t, \hat{p}_t)$, with (\ref{eq:MILP_BESS_charge_discharge_cst1})-(\ref{eq:MILP_BESS_charge_discharge_cst2}) relaxed, provides the dual variables $\phi_t$ and the following objective 
\begin{align}\label{eq:dispatch_dual_objective}
	G \big(x_t, \hat{p}_t, \phi_t \big) = & \sum_{t\in \mathcal{T}} \bigg[ \phi^{cha}_t S^c + \phi^{dis}_t S^d  - \phi^{S^{min}}_t S^{min}  \notag \\
	&  + \phi^{S^{max}}_t S^{max}- \phi^{Y^{min}}_t  Y^{min} + \phi^{Y^{max}}_t  Y^{max}  \notag\\
	& + \phi^{S^i}  S^i +   \phi^{S^f} S^f  - \phi^{\delta x^-}_t (x_t - p P_c) \notag \\
	&  + \phi^{\delta x^+}_t (x_t + p P_c)  + \phi^{y^G}_t \hat{p}_t \bigg]  . 
\end{align}
Then, the dual of the dispatch problem $\min_{y_t \in \Omega(x_t, \hat{p}_t)}  J \big(x_t, y_t \big)$ is
\begin{align}\label{eq:dispatch_dual}
\max_{\boldsymbol{\phi_t \in \Phi}} & \ G \big(x_t, \hat{p}_t, \phi_t \big) ,
\end{align}
with the set of constraints $\Phi$ defined by
\begin{subequations}\label{eq:dispatch_dual_cst}
	\begin{align}	
	& \phi^y_t - \phi^{Y^{min}}_t +  \phi^{Y^{max}}_t - \phi^{\delta x ^-}_t +  \phi^{\delta x ^+}_t =  - \pi_t \Delta t  , \notag \\
	&  \forallt  \quad [y_t] \label{eq:dispatch_dual_cst1} \\
	& - \phi^{\delta x^-}_t \leq \beta \pi_t \Delta t    ,  \ \forallt \quad [\delta x^-_t] \label{eq:dispatch_dual_cst2}  \\
	&- \phi^{\delta x^+}_t  \leq \beta \pi_t \Delta t  ,  \ \forallt \quad [\delta x^+_t] \label{eq:dispatch_dual_cst3}  \\
	& \phi^{dis}_1 - \phi^{y}_1 + \phi^{S^i} \frac{\Delta t}{\eta^d} \leq 0   \label{eq:dispatch_dual_cst4}  \quad [y^{dis}_1] \\
	& \phi^{dis}_t - \phi^{y}_t + \phi^{y^s}_t \frac{\Delta t}{\eta^d} \leq 0 , \ \forallt \setminus \{1\}   \label{eq:dispatch_dual_cst5} \quad [y^{dis}_t] \\
	& \phi^{cha}_1 + \phi^{y}_1 - \phi^{S^i} \eta^c \Delta t   \leq 0   \label{eq:dispatch_dual_cst6} \quad [y^{cha}_1] \\
	& \phi^{cha}_t + \phi^{y}_t - \phi^{y^s}_t \eta^c \Delta t \leq 0 , \ \forallt \setminus \{1\}   \label{eq:dispatch_dual_cst7}  \quad [y^{cha}_t] \\	
	& - \phi^{S^{min}}_1 + \phi^{S^{max}}_1 + \phi^{S^i} - \phi^{y^s}_2 \leq 0  \label{eq:dispatch_dual_cst8} \quad [y^s_1] \\
	& - \phi^{S^{min}}_t + \phi^{S^{max}}_t + \phi^{y^s}_{t-1} - \phi^{y^s}_t \leq 0  , \notag \\
	&  \forallt \setminus \{1, 2, T\}  \label{eq:dispatch_dual_cst9} \quad [y^s_t]\\
	& - \phi^{S^{min}}_{T} + \phi^{S^{max}}_{T} + \phi^{S^f}  + \phi^{y^s}_{T} \leq 0  \label{eq:dispatch_dual_cst10} \quad [y^s_{T}] \\
	& 	- \phi^{y}_t + \phi^{y^G}_t \leq 0 , \ \forallt, \label{eq:dispatch_dual_cst11} \quad [y^{G}_t].
	\end{align}
\end{subequations}
The worst-case dispatch problem $ \max_{\hat{p}_t \in \mathcal{P}} \big[\min_{y_t \in \Omega(x_t, \hat{p}_t)}  J \big(x_t, y_t \big) \big] $ is equivalent to 
\begin{align}\label{eq:worst_case_dispatch}
R(x_t) &  = \max_{\hat{p}_t \in \mathcal{P}, \ \phi_t \in \Phi}  \ G \big(x_t, \hat{p}_t, \phi_t \big) .
\end{align}
Overall, (\ref{eq:RO_min_max_min}) becomes a min-max problem
\begin{align}\label{eq:RO_min_max_final}	
\min_{x_t \in \mathcal{X}}  \bigg[ \max_{\hat{p}_t \in \mathcal{P}, \ \phi_t \in \Phi} & \ G \big(x_t, \hat{p}_t, \phi_t \big)\bigg]  ,
\end{align}
that can be solved using a Benders decomposition technique such as BD or CCG, between a master problem, that is linear, and a sub-problem, that is bilinear, since 
$G$ has the terms $\phi^{y^G}_t \hat{p}_t = \phi^{y^G}_t  \hat{p}^{(0.5)}_t - \phi^{y^G}_t z_t p_t^{min}$. 
It is possible to linearize the products of the binary and continuous variables $z_t \phi^{y^G}_t$ of $G$ by using a standard integer algebra trick \cite{savelli2018new} with the following constraints $\forallt$
\begin{subequations}\label{eq:linearization2}
	\begin{align}		
	- M_t^- z_t& \leq \alpha_t \leq M_t^+ z_t\\
	- M_t^-(1-z_t) & \leq \phi^{y^G}_t-\alpha_t \leq M_t^+(1-z_t) ,
	\end{align}
\end{subequations}
where $M_t^\pm$ are the big-M's values of $\phi^{y^G}_t$ and $\alpha_t$ is an auxiliary continuous variable.
The definition of the uncertainty set (\ref{eq:G_set}) with binary variables, based on the Proposition 2 of \cite{zhao2012robust}, is essential to linearize $G$.

\subsection{Controller formulation}\label{sec:controller}

The controller uses as parameters the engagements $x_t$, the system last measured values, and renewable generation intraday point forecasts. It computes at each period $t$ the set-points from $t$ to the last period $T$ of the day. The formulation is the following MILP
\begin{align}\label{eq:det_compact-controller}
\min_{y_t \in \Omega(x_t, \hat{p}_t)}& \ J \big(x_t, y_t \big)  .
\end{align}

\section{Solution methodology}\label{sec:ro_benders}

Following the methodology described by \cite{bertsimas2012adaptive,jiang2011robust}, a two-level algorithm can be used to solve the two-stage RO problem with a Benders-dual cutting plane algorithm.
The following master problem (MP) is solved iteratively by adding new constraints to cut off the infeasible or non-optimal solutions
\begin{subequations}\label{eq:master_problem}	
	\begin{align}
	\min_{x_t \in \mathcal{X}, \ \theta} &    \ \theta \\
	& \theta \geq G \big(x_t, \alpha_{t,l}, \phi_{t,l} \big),  \quad l = 1 \ldots L \label{eq:optimality_cut} \\
	&  G \big(x_t, \tilde{\alpha}_{t,k}, \tilde{\phi}_{t,k} \big) \leq 0, \quad k = 1 \ldots K , \label{eq:feasibility_cut}
	\end{align}
\end{subequations}
where constraints (\ref{eq:optimality_cut}) represent the optimality cuts, generated by retrieving the optimal values $\alpha_{t,l},  \phi_{t,l}$ of (\ref{eq:worst_case_dispatch}), while constraints (\ref{eq:feasibility_cut}) represent the feasibility cuts, generated by retrieving the extreme rays $\tilde{\alpha}_{t,k}, \tilde{\phi}_{t,k}$ of (\ref{eq:worst_case_dispatch}), and $\theta$ is the optimal value of the second-stage problem.
Section \ref{sec:case_study} also reports results for the CCG algorithm. However, for the sake of the paper's clarity, the CCG implementation is detailed in Appendix \ref{appendix:ccg}.

\subsection{Algorithm convergence}\label{sec:convergence-checking}

First\footnote{
The comments of this subsection apply to the BD and CCG algorithms.}, we make the \textit{relatively complete recourse} assumption that the SP is feasible for any engagement plan $x_t$ and generation trajectory $\hat{p}_t$. This assumption is valid in the capacity firming framework where curtailment is allowed. If the system faces underproduction where $x_t$ is large, the generation is 0, and the BESS discharged, penalties are applied. If it encounters overproduction where $x_t$ is close to 0, the generation is large, and the BESS is charged, the excess of generation is curtailed. In both cases, there is always a feasible dispatch.
Notice, when the relatively complete recourse assumption does not hold, \cite{bertsimas2018scalable} propose an extension of the CCG algorithm.

Second, the convergence of the relaxed SP is checked at each iteration of the algorithm by ensuring there is no simultaneous charge and discharge. However, such a situation should not occur because, in case of overproduction, the excess of generation can be curtailed. Simultaneous charging and discharging could indeed be an equivalent solution to dissipate the excess energy. That solution can be avoided in practice by adding a small penalty for using the storage system. However, we never observed simultaneous charge and discharge over the hundreds of simulations carried out.
Thus, it is not required to implement an extension of the BD or CCG algorithm that handles a linear two-stage robust optimization model with a mixed-integer recourse problem such as proposed by \cite{zhao2012exact}.

Finally, the overall convergence of the algorithm toward the optimal solution is checked. Indeed, depending on the big-M's values, the algorithm may converge by reducing the gap between the MP and SP. However, it does not ensure an optimal solution. Therefore, once the convergence between the MP and SP is reached at iteration $j=J$, the objective of the MP at $J$ is compared to the objective of the MILP formulation (\ref{eq:det_compact}) using the worst-case generation trajectory $\hat{p}_t^{\star,J}$ as parameters. If the absolute gap $|MILP^J - MP^J |$ is higher than a convergence threshold $\epsilon$, the convergence is not reached. Then, larger big-M's values are set, and the algorithm is restarted until convergence or a stopping criterion is reached.

\subsection{Benders-dual cutting plane algorithm}

Figure \ref{algo_benders} depicts the Benders-dual cutting plane algorithm implemented.
The initialization step consists of setting the initial big-M's values $M_t^- = 1$ and $M_t^+ = 0 \ \forallt$, the time limit resolution of the sub-problem (\ref{eq:worst_case_dispatch}) to 10 s, and the threshold convergence $\epsilon $ to 0.5 \euro. Let $MP^j $, $SP^j$, be the MP and SP objective values at iteration $j$, the lower and upper bounds, respectively, and $MILP^J$ the MILP objective value using the worst renewable generation trajectory $\hat{p}_t^{\star,J}$ at iteration $J$.
Note: the big-M's values cannot exceed 500 to maintain the computation time to a few minutes. The BD algorithm has never reached this limit for all the simulations, which is not the case with the CCG as depicted in Table \ref{tab:BD_vs_CCG} and explained in Section \ref{sec:BD_vs_CCG}.
\begin{figure}
	\begin{algorithmic}
		\STATE Initialization.
		\STATE Warm-start: build the initial set of cuts $\{\theta_i\}_{1\leq i \leq I}$.
		\WHILE{$|MILP^J - MP_2^J |>  \epsilon$ and $M_t^- <500$ }
		\STATE Initialize $j=0$, solve the MP (\ref{eq:master_problem}) and retrieve $x_{t,0}$.
		\WHILE{the 10 last $|MP^j - SP^j| $ are not $< \epsilon$ }
		\STATE Solve the SP (\ref{eq:worst_case_dispatch}) with $x_{t,j}$ as parameters:
		\IF{the SP is unbounded}
		\STATE Retrieve the extreme rays $ \tilde{\alpha}_{t,k}, \tilde{\phi}_{t,k}, \ 0 \leq k \leq j$.
		\STATE Add the $k$-th feasibility cut: $ G \big(x_{t,j},\tilde{\alpha}_{t,k}, \tilde{\phi}_{t,k} \big) \leq 0$.
		\ELSE 
		\STATE Retrieve the optimal values $\alpha_{t,l}, \phi_{t,l}, \ 0 \leq l \leq j$.
		\STATE Add the $l$-th optimality cut: $ \theta \geq G \big(x_{t,j}, \alpha_{t,l}, \phi_{t,l} \big)$.
		\STATE Update the upper bound $SP^j= R(x_{t,j})$.
		\STATE SP check: no simultaneous charge and discharge.
		\ENDIF
		\STATE Solve the MP (\ref{eq:master_problem}): get the optimal values $\theta_j, x_{t,j}$.
		\STATE Update the lower bound $MP^j=\theta_j$ and  $j=j+1$.
		\ENDWHILE
		\STATE $j=J$: convergence between the SP and MP is reached. Check convergence with MILP: get $\hat{p}_t^{\star,J}$ from $SP^J$ and compute $MILP^J$ (\ref{eq:det_compact}).
		\IF{$|MILP^J - MP^J |>  \epsilon$}
		\IF{$M_t^- \leq 50$}
		\STATE Update big-M's values $M_t^- = 10 + M_t^- \ \forallt$. 
		\ELSE
		\STATE Update big-M's values $M_t^- = 100 + M_t^- \ \forallt$.
		\ENDIF
		\STATE Reset $j$ to 0 and restart algorithm with a new $MP$.
		\ENDIF
		\ENDWHILE
		\STATE Retrieve the final $x_{t,J}$ engagement.
	\end{algorithmic}
	\caption{Benders-dual cutting plane algorithm.}\label{algo_benders}
\end{figure}

\section{Case Study}\label{sec:case_study}

The BD and CCG algorithms are compared on the ULi\`ege case study. It comprises a PV generation plant with an installed capacity $P_c =$ 466.4 kWp. The PV generation is monitored on a minute basis, and the data are resampled to 15 minutes. The dataset contains 350 days from August 2019 to November 2020, missing data during March 2020.
The NFs approach is compared to a widely used neural architecture, referred to as Long Short-Term Memory (LSTM). In total, eight versions of the planner are considered. Four RO versions: BD-LSTM, BD-NF, CCG-LSTM, and CCG-LSTM.
Four deterministic versions: the oracle that uses perfect knowledge of the future, a benchmark that uses PV nominal point forecasts, and two versions using NFs and LSTM PV quantiles. The set of PV quantiles is $\mathcal{Q} =  \{q =10 \%, \ldots, 50 \%\}$.
The controller uses PV intraday point forecasts and the day-ahead engagements computed by the planners to compute the set-points and the profits. They are normalized by the profit obtained with the oracle planner and expressed in \%.

Section \ref{sec:numerical_settings} presents the numerical settings. Section \ref{sec:res-static-ro} provides the results of the sensitivity analysis for several risk-averse pairs $[p_t^{min}=\hat{p}^{(q)} , \Gamma]$, with $q = 10, \ldots, 40 \%$, and $\Gamma = 12$, 24, 36, 48. Section \ref{sec:res-syn-ro} investigates the dynamic risk-averse parameter selection. Finally, Section \ref{sec:BD_vs_CCG} compares the BD and CCG algorithms.

\subsection{Numerical settings}\label{sec:numerical_settings}

The testing set is composed of thirty days randomly selected from the dataset.
The simulation parameters of the planners and the controller are identical. The planning and controlling periods duration are $\Delta t = 15$ minutes. The peak hours are set between 7 pm and 9 pm (UTC+0). 
The ramping power constraint on the engagements are $\Delta X_t = 7.5 \% P_c$ ($15 \% P_c$) during off-peak (peak) hours. The lower bounds on the engagement $X_t^{min}$ and the net power $Y_t^{min}$ are set to 0 kW. The upper bound on the engagement $X_t^{max}$ and the net power $Y_t^{max}$ are set to $P_c$. Finally, the engagement tolerance is $ p P_c = 1 \% P_c$, and the penalty factor $\beta = 5$.
%
The BESS minimum $S^{min}$ and maximum capacity are 0 kWh and 466.4 kWh, respectively. It is assumed to be capable of fully charging or discharging in one hour $S^d = S^c = S^{max} / 1$ with charging and discharging efficiencies $\eta^d = \eta^c = 95$ \%. Each simulation day is independent with a fully discharged battery at the first and last period $S^i = S^f = 0$ kWh. 
%
The Python Gurobi library is used to implement the algorithms in Python 3.7, and Gurobi\footnote{\url{https://www.gurobi.com/}} 9.0.2 to solve all the optimization problems. Numerical experiments are performed on an Intel Core i7-8700 3.20 GHz based computer with 12 threads and 32 GB of RAM running on Ubuntu 18.04 LTS. 

\begin{figure}[htbp]
	\centering
	\begin{subfigure}{.25\textwidth}
		\centering
		\includegraphics[width=\linewidth]{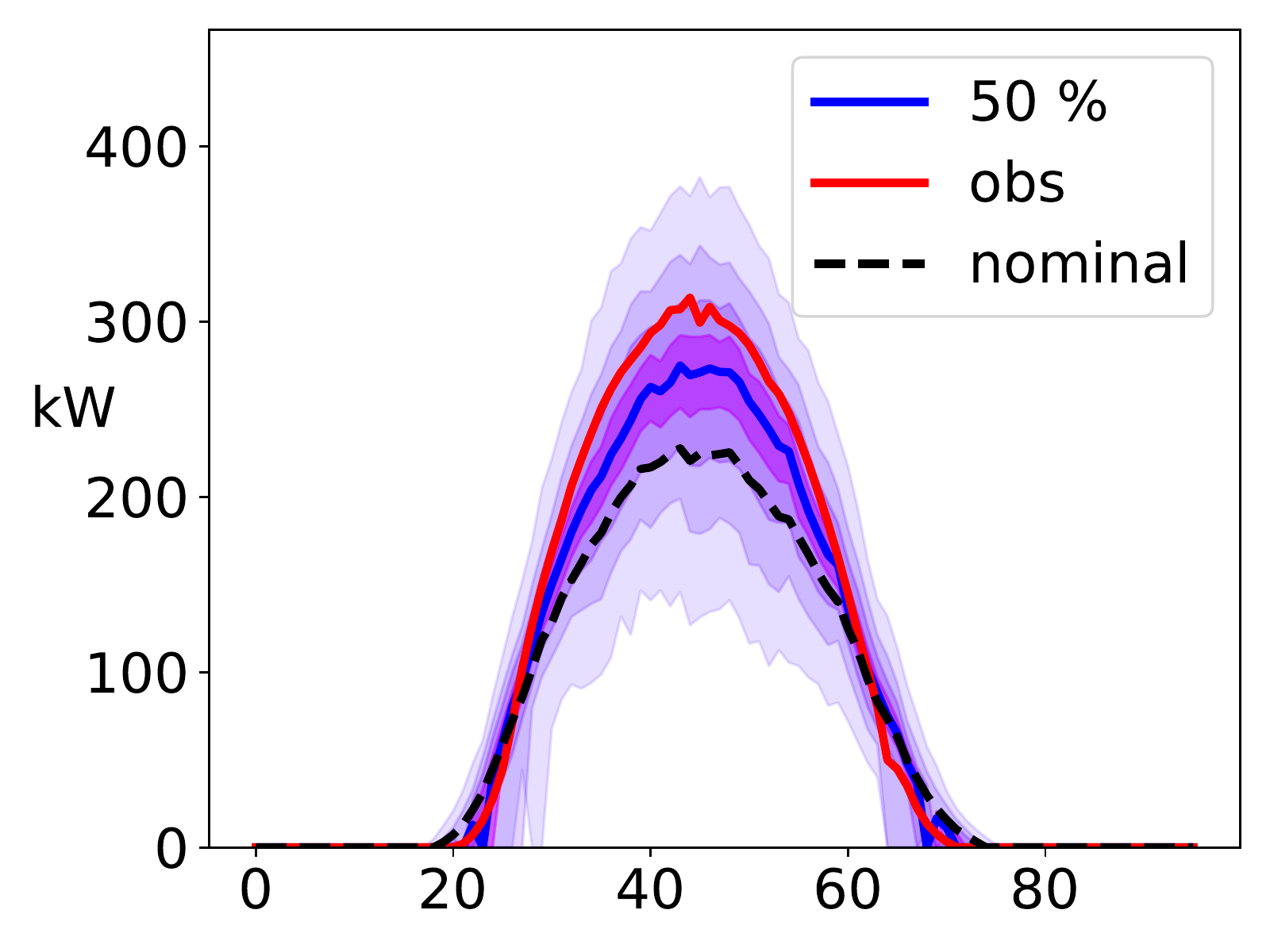}
		\caption{LSTM PV quantile forecasts.}
		\label{fig:quantile_forecast_LSTM}
	\end{subfigure}%
	\begin{subfigure}{.25\textwidth}
		\centering
		\includegraphics[width=\linewidth]{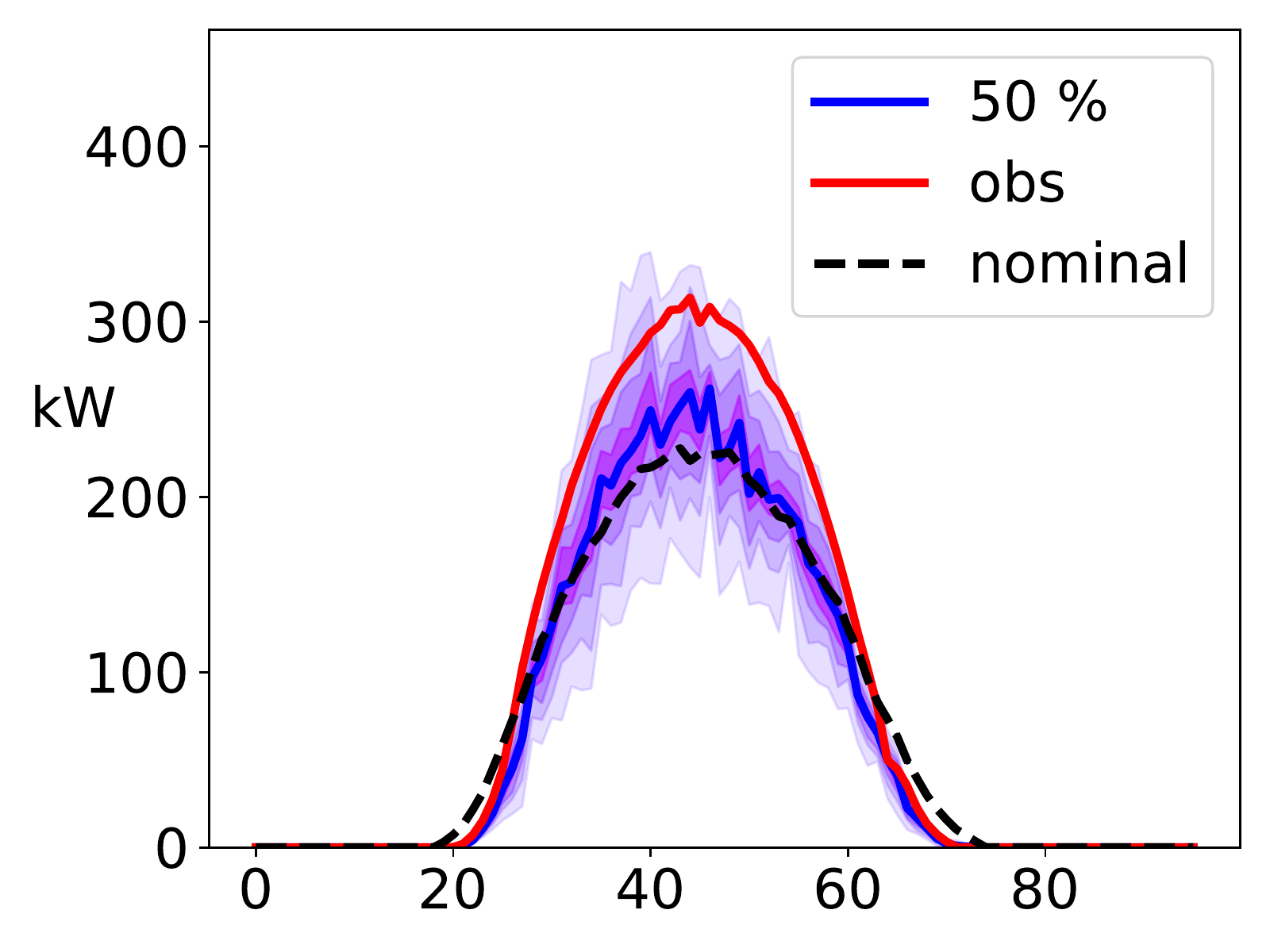}
		\caption{NFs PV quantile forecasts.}
		\label{fig:quantile_forecast_NF}
	\end{subfigure}	
	\begin{subfigure}{.25\textwidth}
		\centering
		\includegraphics[width=\linewidth]{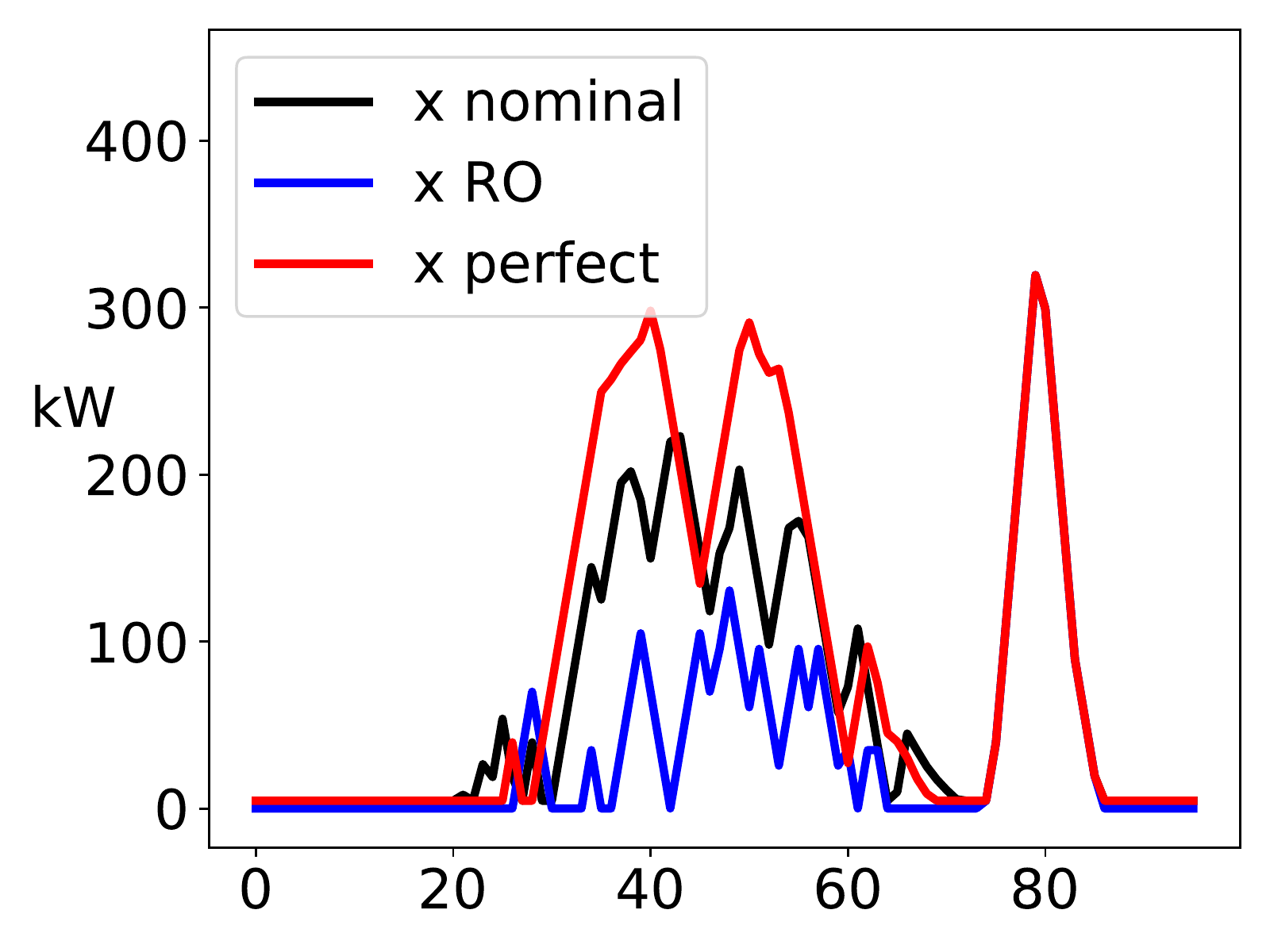}
		\caption{Engagement plan.}
		\label{fig:x}
	\end{subfigure}%
	\begin{subfigure}{.25\textwidth}
		\centering
		\includegraphics[width=\linewidth]{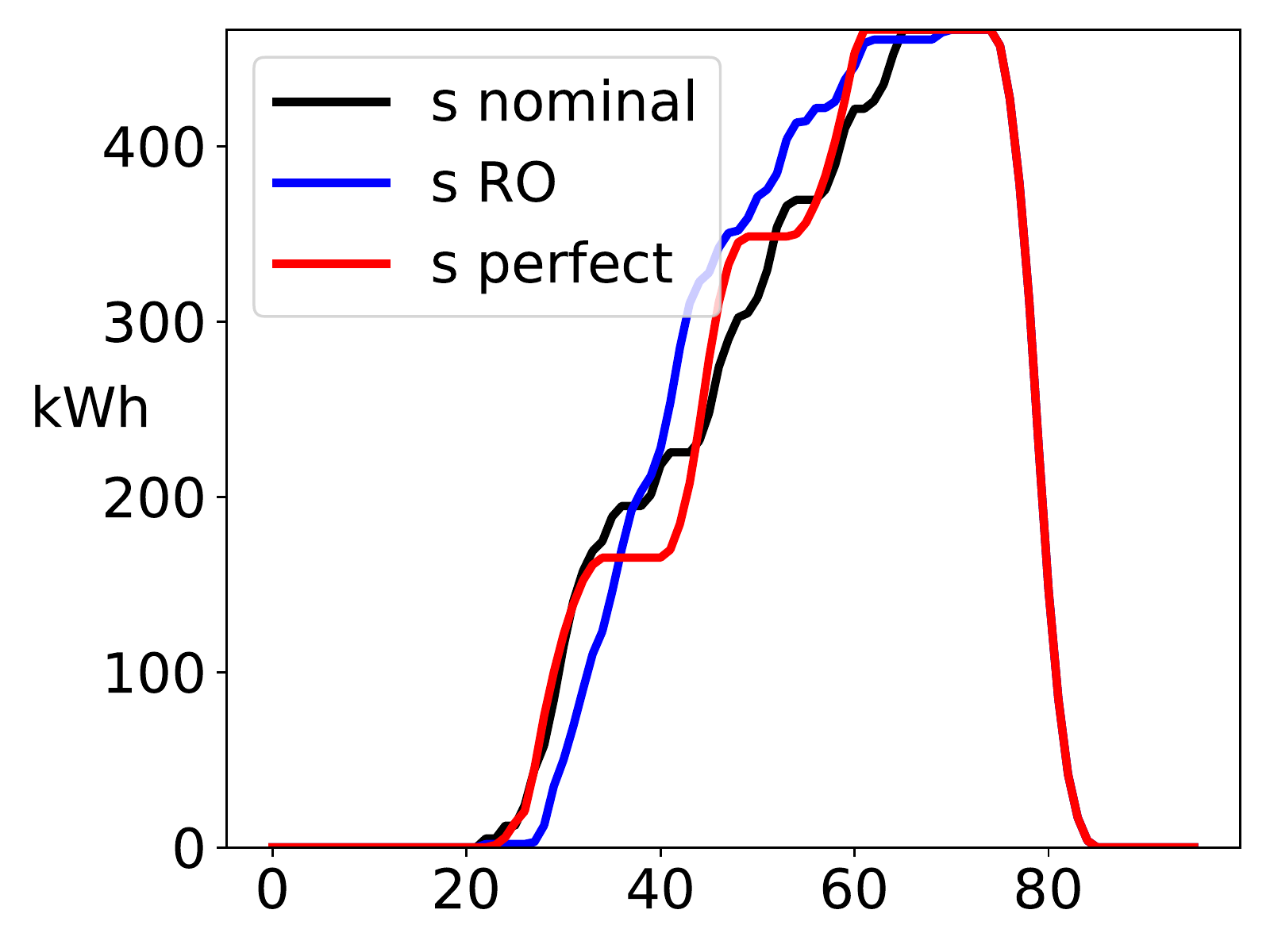}
		\caption{BESS state of charge.}
		\label{fig:s}
	\end{subfigure}
	\caption{Illustration of the results on $\convdate$.}
	\label{fig:x_s_res}
\end{figure}

Figures \ref{fig:quantile_forecast_LSTM} and \ref{fig:quantile_forecast_NF} illustrate the LSTM and NFs PV quantile forecasts, observation, and nominal point forecasts on $\convdate$. Figures \ref{fig:x} and \ref{fig:s} provide the engagement plan (x) and the BESS state of charge (s) computed with the BD-RO planner, the deterministic planner with the nominal point forecasts, and the perfect knowledge of the future.

\subsection{Constant risk-averse parameters strategy}\label{sec:res-static-ro}

\begin{figure}[tb]
	\centering
	\begin{subfigure}{.25\textwidth}
	\centering
	\includegraphics[width=\linewidth]{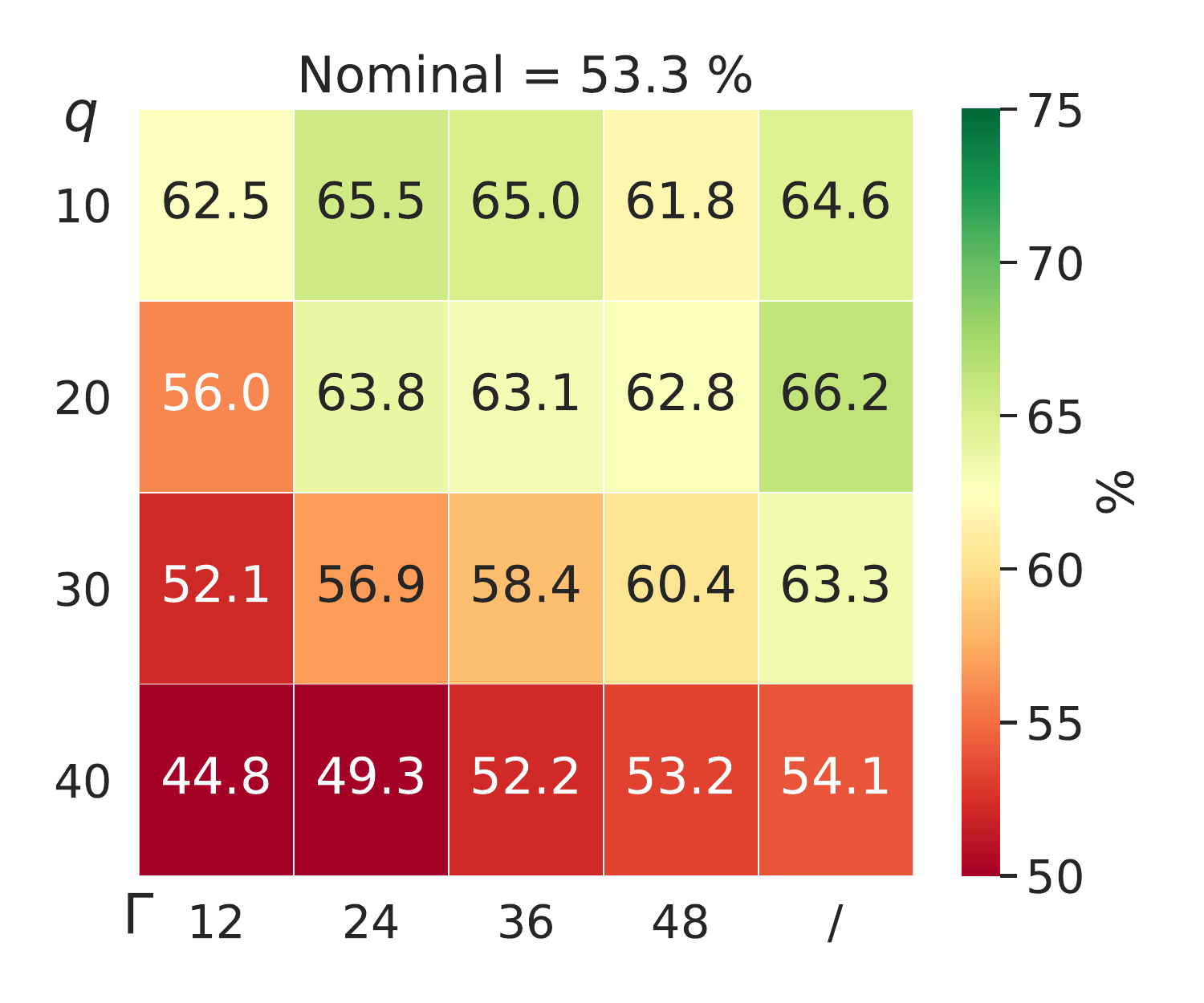}
	\caption{BD-LSTM.}
	\end{subfigure}%
	\begin{subfigure}{.25\textwidth}
		\centering
	\includegraphics[width=\linewidth]{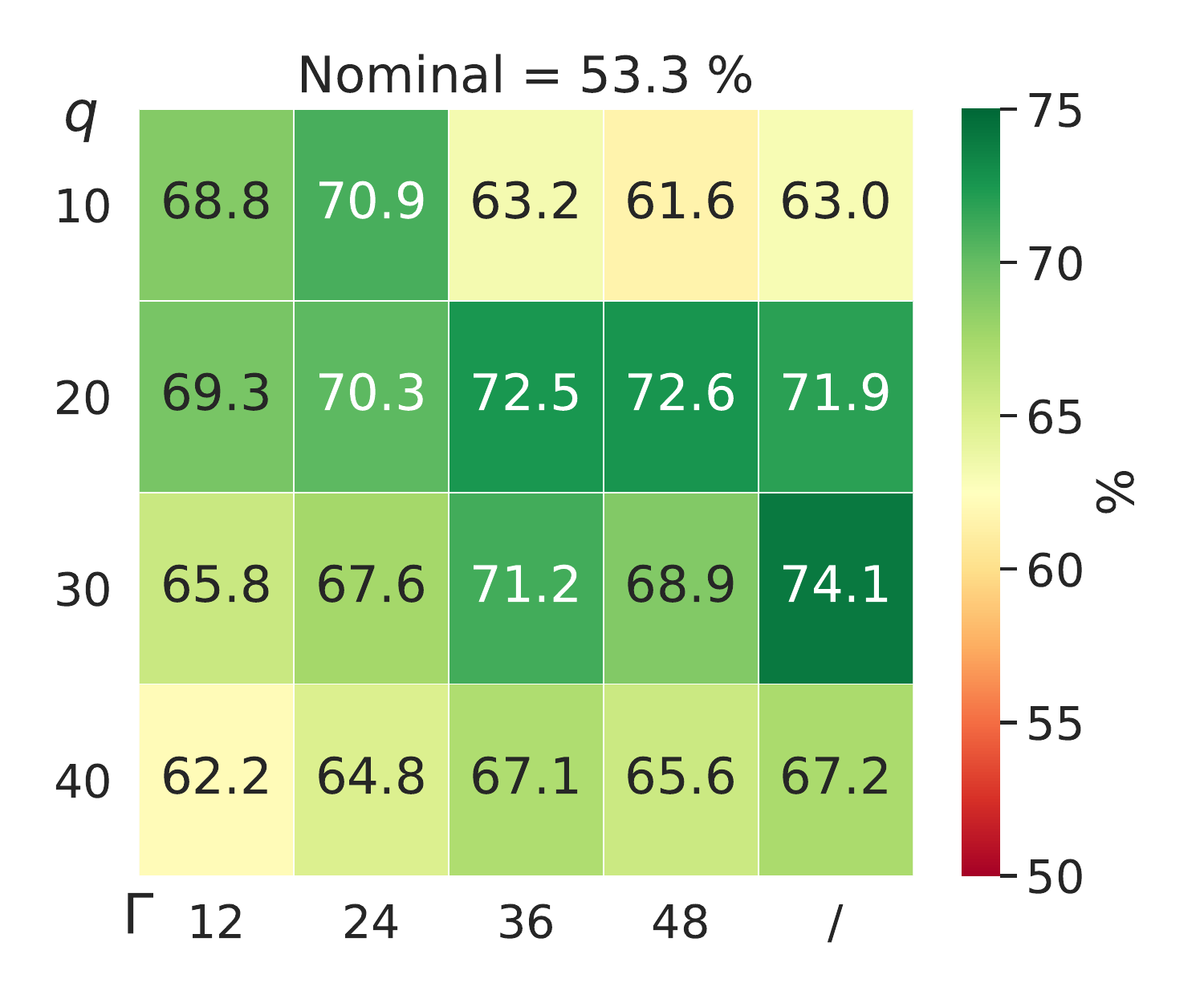}
	\caption{BD-NF.}
	\end{subfigure}
	 \begin{subfigure}{.25\textwidth}
		\centering
	\includegraphics[width=\linewidth]{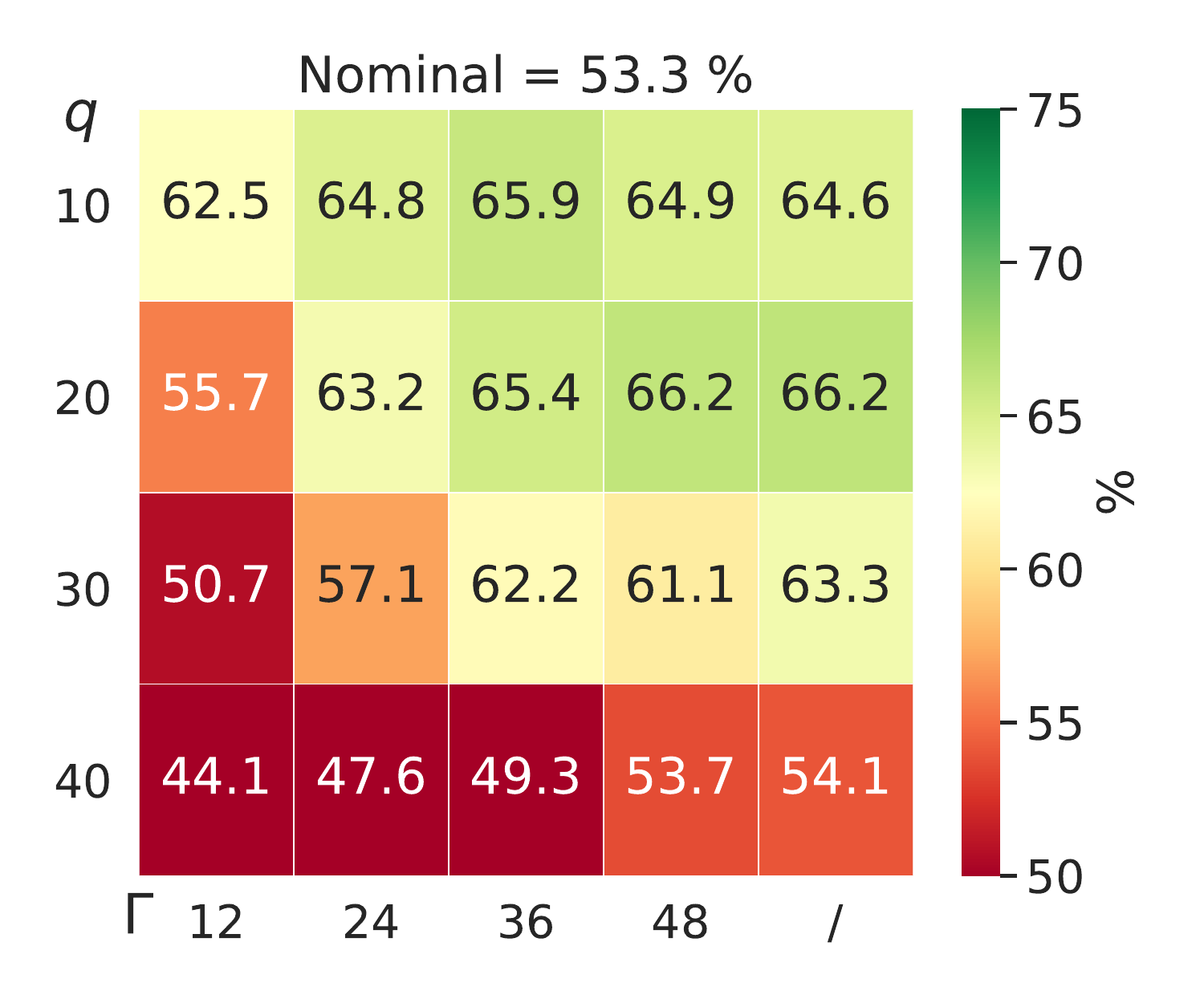}
	\caption{CCG-LSTM.}
	\end{subfigure}%
	\begin{subfigure}{.25\textwidth}
		\centering
	\includegraphics[width=\linewidth]{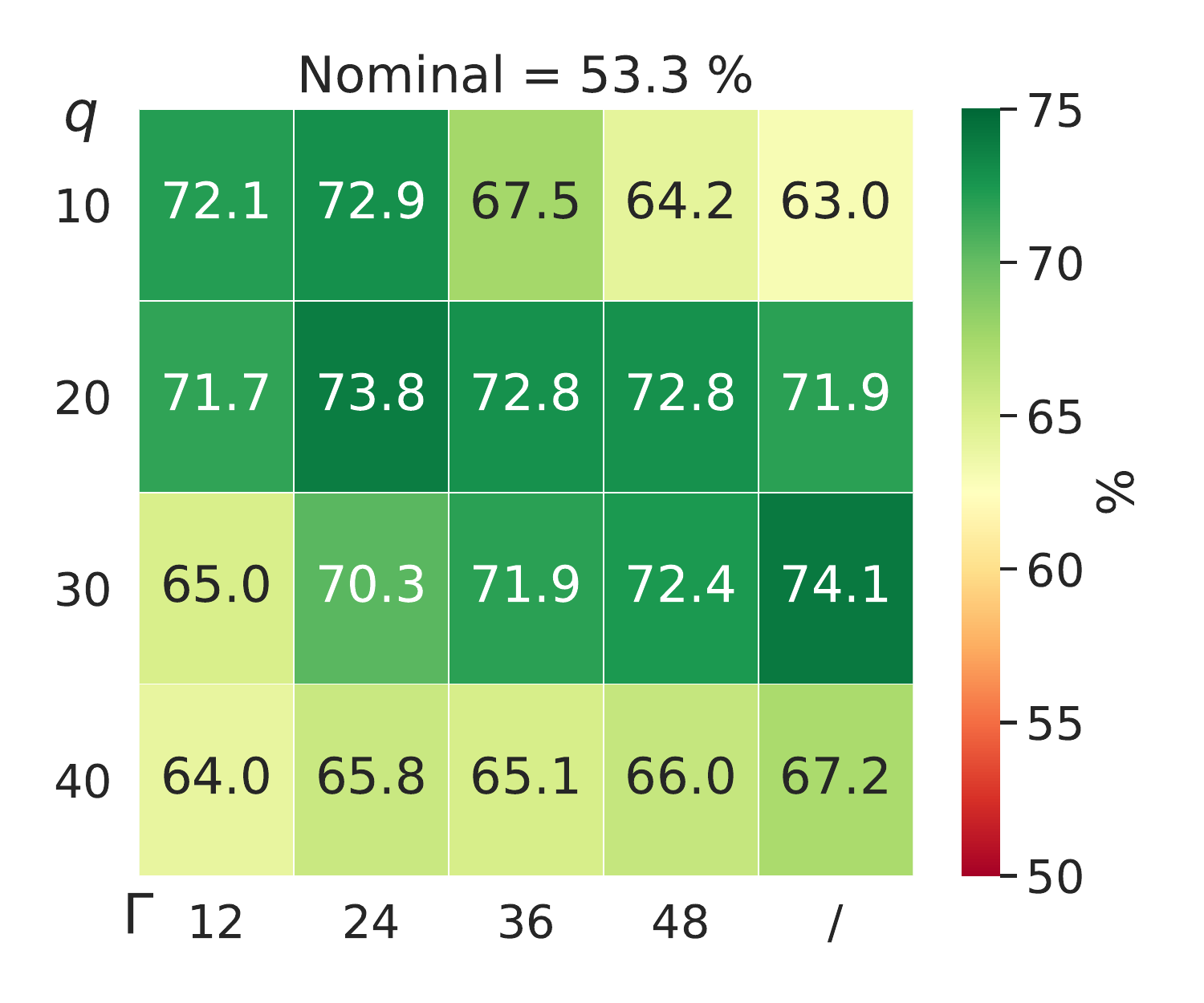}
	\caption{CCG-NF.}
	\end{subfigure}
	\caption{Results with constant risk-averse parameters. Normalized profit (\%) of the BD and CCG RO planners ($[\Gamma, q]$), deterministic ($[/,q]$) planner, and the reference that is the deterministic planner with point-forecasts (Nominal). Left part: LSTM quantiles, right part: NF quantiles.}
	\label{fig:statis-ro-res}
\end{figure}

The risk-averse parameters of the RO approach $[\hat{p}^{(q)}, \Gamma]$ are constant over the dataset. One way to identify the optimal pair is to perform a sensitivity analysis \cite{wang2015robust}. 
Figure \ref{fig:statis-ro-res} provides the normalized profits of the BD-RO, CCG-RO and deterministic planners using PV quantiles, left with LSTM and right with NFs, and nominal point forecasts. 
The RO and deterministic planners outperform by a large margin the baseline. The latter, the deterministic planner with nominal point forecasts, cannot deal with PV uncertainty and achieved only 53.3 \%. Then, the planners using NFs quantiles significantly outperform the planners with LSTM quantiles. 
Overall, the CCG algorithm achieved better results for almost all pairs of risk-averse parameters.
The highest profits achieved by the CCG-NF, BD-NF and NF-deterministic planners are 73.8 \%, 72.6 \% and 74.1 \%, respectively, with the risk-averse parameters $[q = 20 \%, \Gamma = 24]$, $[q = 20 \%, \Gamma = 48]$, and the quantile 30 \%. 
It should be possible to improve the RO results by tuning the risk-averse parameters $[\hat{p}^{(q)} , \Gamma]$. However, these results emphasize the interest in considering a deterministic planner with the relevant PV quantile as point forecasts that are easy to implement, fast to compute (a few seconds), and less prone to convergence issues than the two-stage RO approach.

\subsection{Dynamic risk-averse parameters strategy}\label{sec:res-syn-ro}

In this section, the risk-averse parameters $[p_t^{min} , \Gamma]$ of the RO approach are dynamically set based on the day-ahead quantile forecasts distribution, and $p_t^{min}$ is not necessarily equal to the same quantile $\hat{p}^{(q)} \ \forallt$.
The motivation of this strategy is to assume that the sharper the quantile forecast distribution around the median is, the more risk-averse the RO approach should be.

Two parameters are designed to this end: (1) the PV uncertainty set max depth $d_q$ to control $p_t^{min}$; (2) the budget depth $d_\Gamma$ to control $\Gamma$. $d_q$ is a percentage of the distance between the median and the 10 \% quantile $d_{50-10}$, and $d_\Gamma$  is a percentage of the total installed capacity $P_c$. Then, two rules are designed to dynamically set the risk-averse parameters $[p_t^{min} , \Gamma]$ for each day of the dataset.
%
For a given day, and the set of time periods where the PV median is non null, the distances between the PV median and the PV quantiles 20, 30, and 40 \% are computed: $d_{50-20}$, $d_{50-30}$, $d_{50-40}$. $p_t^{min}$ is dynamically set at each time period $t$ as follows
\begin{equation}\label{eq:min_rule}
p_t^{min} =  
\begin{cases} 
\hat{p}^{(0.1)}_t   & \text{if } d^{50-20/30/40}_t  > d_q  d^{50-10}_t \\
\hat{p}^{(0.2)}_t   & \text{if } d^{50-20/30}_t  > d_q  d^{50-10}_t  \\
\hat{p}^{(0.3)}_t   & \text{if } d^{50-20}_t  > d_q  d^{50-10}_t  \\
\hat{p}^{(0.4)}_t    & \text{otherwise}  
\end{cases} .
\end{equation}
For a given day, the budget of uncertainty $\Gamma$ is dynamically set based on the following rule
\begin{align}\label{eq:gamma_rule}	
\Gamma =   \#  \{t :  d^{50-10}_t>  d_\Gamma P_c \} .
\end{align}

\begin{figure}[tb]
	\centering
	\begin{subfigure}{.25\textwidth}
		\centering
		\includegraphics[width=\linewidth]{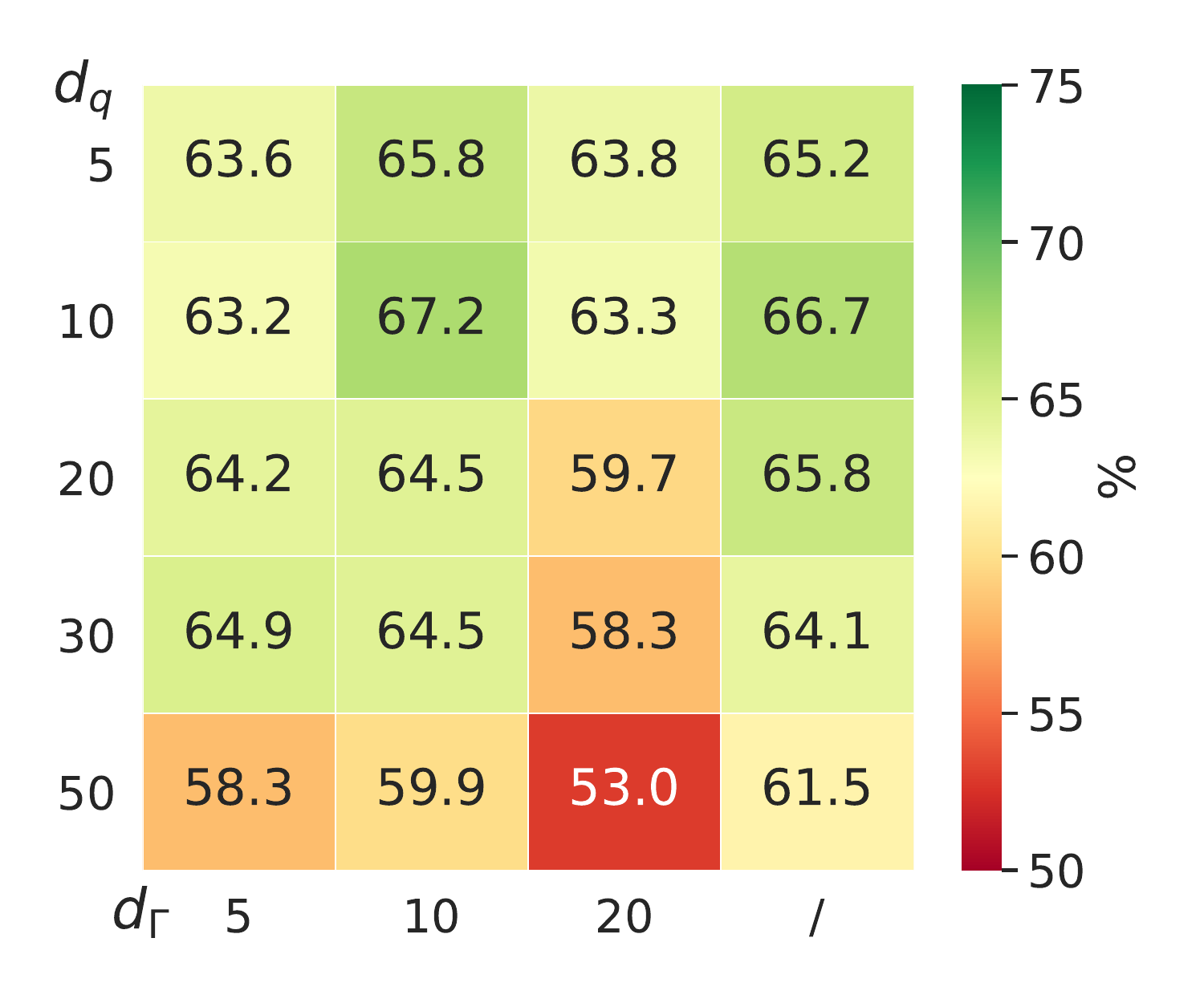}
		\caption{BD-LSTM.}
	\end{subfigure}%
	\begin{subfigure}{.25\textwidth}
		\centering
		\includegraphics[width=\linewidth]{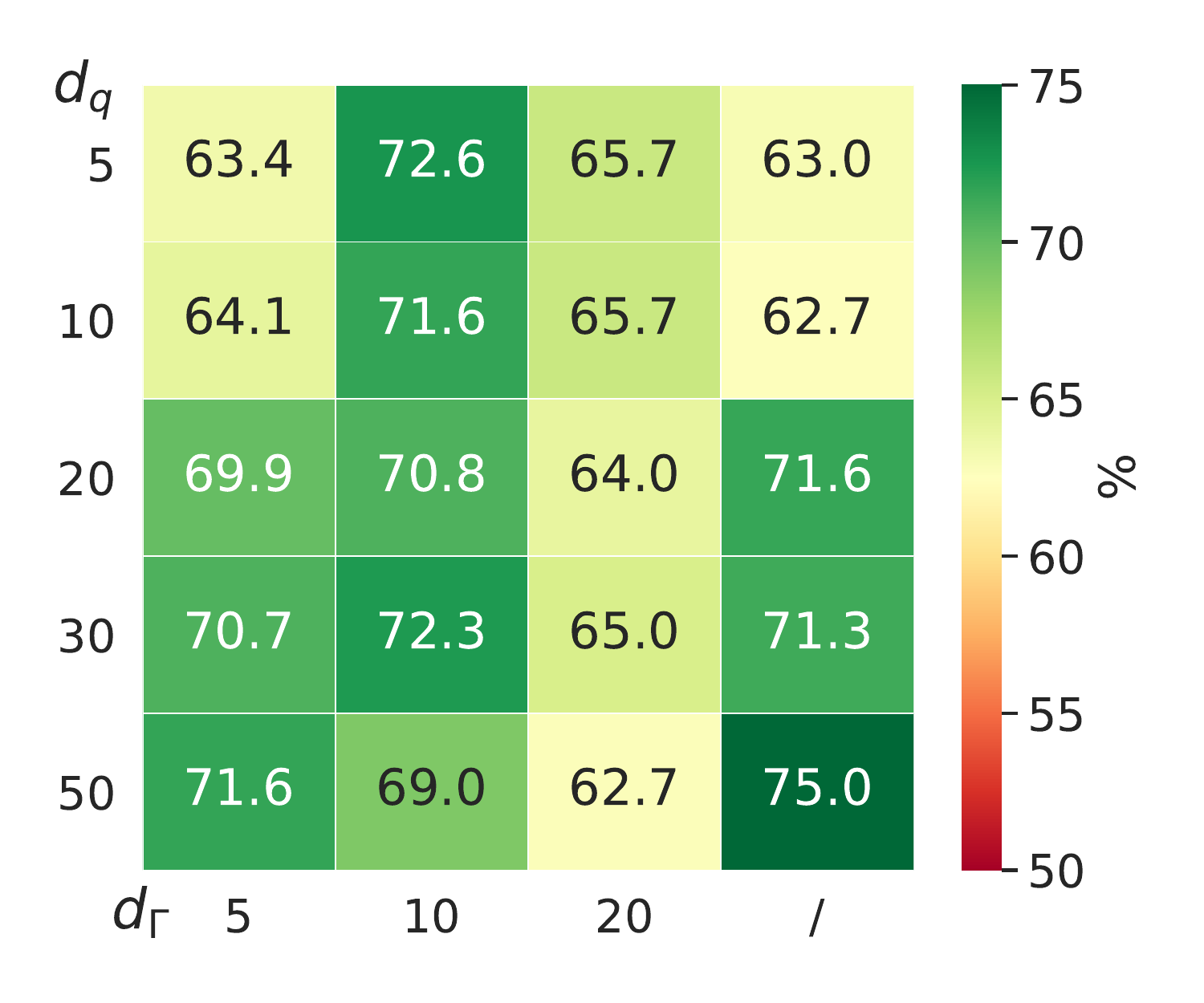}
		\caption{BD-NF.}
	\end{subfigure}
	 	\begin{subfigure}{.25\textwidth}
		\centering
		\includegraphics[width=\linewidth]{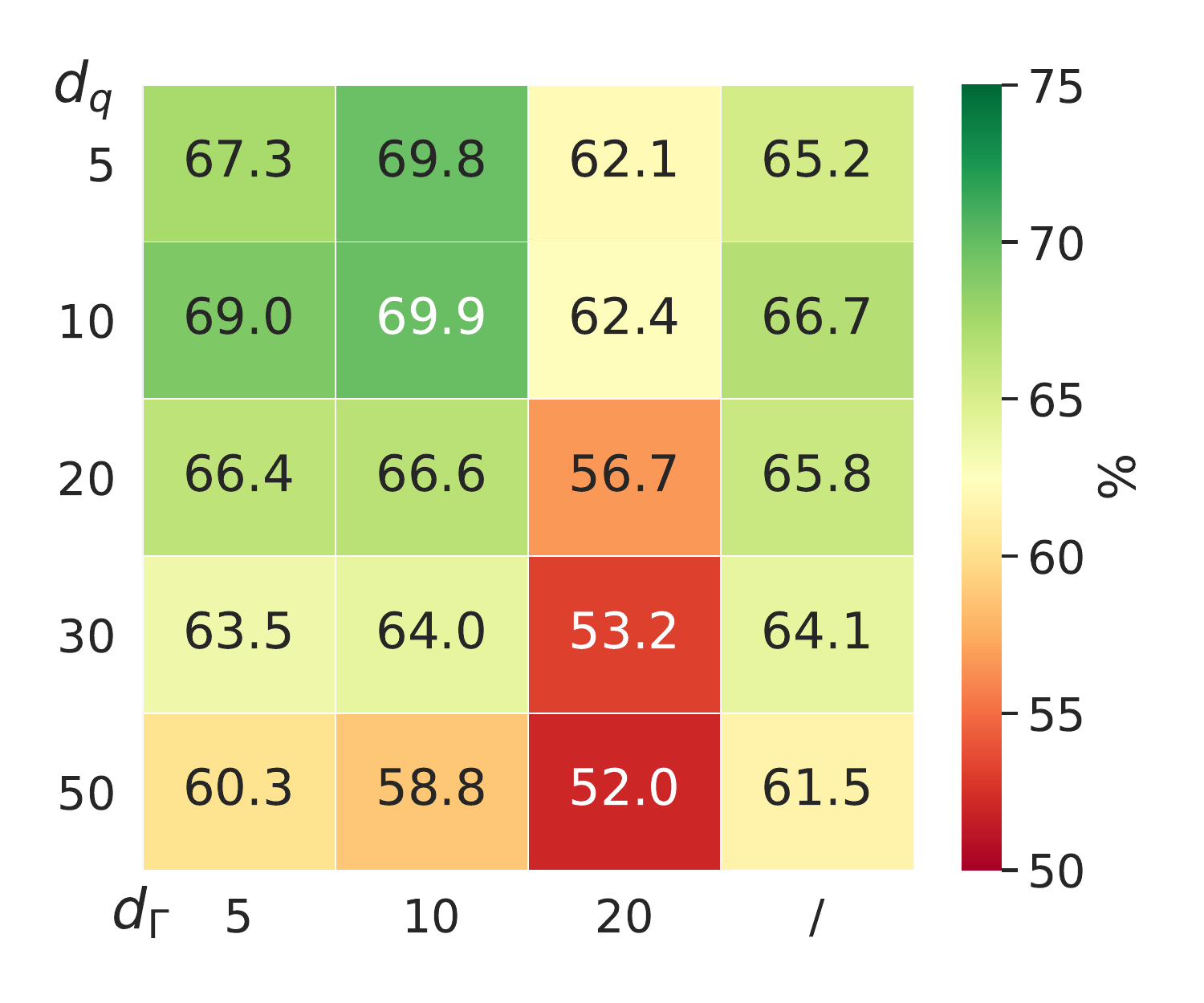}
		\caption{CCG-LSTM.}
	\end{subfigure}%
	\begin{subfigure}{.25\textwidth}
		\centering
		\includegraphics[width=\linewidth]{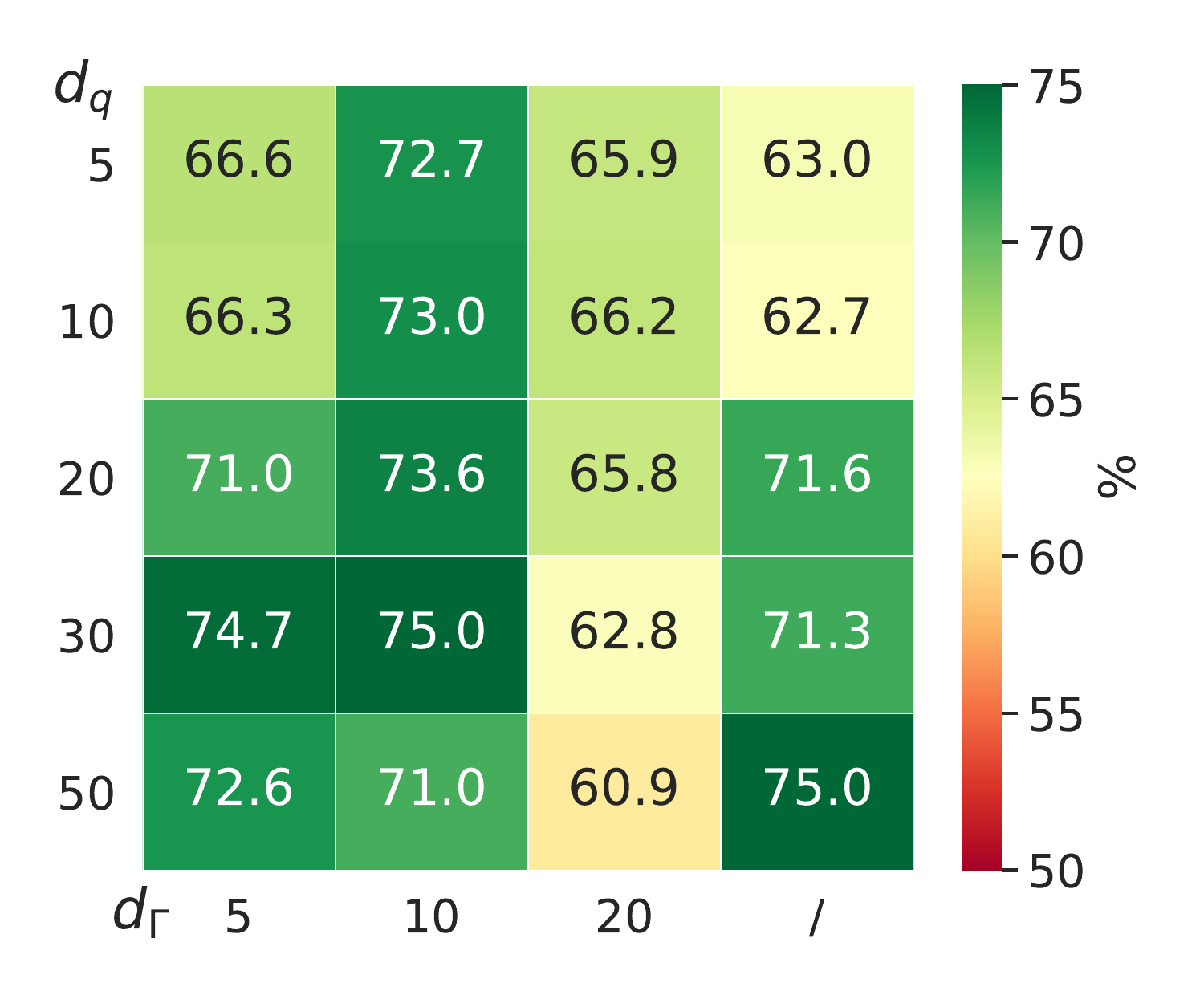}
		\caption{CCG-NF.}
	\end{subfigure}
	\caption{Results with dynamic risk-averse parameters. Normalized profit (\%) of the BD and CCG RO planners ($[d_\Gamma, d_q]$), and deterministic ($[/,d_q]$) planner. Left part: LSTM quantiles, right part: NF quantiles.}
	\label{fig:dyn-ro-res}
\end{figure}

Figure \ref{fig:dyn-ro-res} provides the normalized profits of the CCG-RO, BD-RO, and deterministic planners for several pairs $[d_\Gamma, d_q]$ using both the LSTM and NF quantiles. 
The planners achieved better results when using the NF quantiles. Overall, the results are improved compared to fixed risk-averse parameters for all the planners. 
The highest profits achieved by the CCG-NF, BD-NF and NF-deterministic planners are 75.0 \%, 72.6 \% and 75.0 \%, respectively, with $[d_\Gamma, d_q] = [10, 30]$, $[d_\Gamma, d_q] = [10, 5]$, and $d_q = 50 \%$.

\subsection{BD and CCG comparison}\label{sec:BD_vs_CCG}

Table \ref{tab:BD_vs_CCG} provides a comparison of the BD and CCG algorithms when using NF quantiles for both the static and dynamic robust optimization strategies.
Overall, the CCG algorithm converges in 5-10 iterations instead of 50-100 for BD. Therefore, the CCG computes the day-ahead planning in approximately 10 seconds, ten times faster than BD. This observation is consistent with \cite{zeng2013solving} that demonstrated the CCG algorithm converges faster than BD.
Let $n$ be the number of extreme points of the uncertainty set $\mathcal{P}$ and $m$ of the space $\Phi$ defined by constraint (\ref{eq:dispatch_dual_cst}). The BD algorithm computes an optimal solution in $O(nm)$ iterations, and the CCG procedure in $O(n)$ iterations \cite{zeng2013solving}.
Note: the BD algorithm is still competitive in an operational framework as it takes on average 1-2 minutes to compute the day-ahead planning.
However, we observed that the CCG does not always converge to an optimal solution (see Section \ref{sec:convergence-checking}), which never happened with the BD algorithm. Fortunately, these cases amount to only a few \% of the total instances. 
Overall, the CCG algorithm achieved better results than the BD for almost all the risk-averse parameters.
Finally, in our opinion, both algorithms require the same amount of knowledge to be implemented. Indeed, the only difference is the MP as the SP are solved identically. 
\begin{table}[tb]
\renewcommand{\arraystretch}{1.25}
	\begin{center}
		\begin{tabular}{lrrrr}
			\hline \hline
Algorithm & RO-type & $\overline{t}$ & $1_\%$ & $J^{max}$  \\ \hline
BD-NF   & static  & 85.2 (151.9)   & 0.0 & 72.6 \\
CCG-NF  & static  & 7.5 (6.0)      & 1.9 & 73.8  \\ \hline
BD-NF   & dynamic & 102.3 (107.3)  & 0.0 & 72.6 \\
CCG-NF  & dynamic & 9.2 (5.5)      & 4.2 & 75.0 \\ \hline\hline
\end{tabular}
\caption{BD vs CCG statistics. \\
$\overline{t}$ (s) is the averaged computation time per day with the standard deviation in the bracket. $1_{\%}$ (\%) is the \% of instances that did not terminate with optimality. $\overline{t}$ and $1_{\%}$ are computed over all days of the testing set and for all pair of constant (dynamic) risk-averse parameters $[p_t^{min} , \Gamma]$ ($[d_\Gamma, d_q]$). $J^{max}$ (\%) is the best-normalized profit achieved using the NF quantiles over all risk-averse parameters.}
\label{tab:BD_vs_CCG}
\end{center}
\end{table}
%

\section{Conclusion}\label{sec:conclusions}

The core contribution of this study is to address the two-phase engagement/control problem in the context of capacity firming. 
A secondary contribution is to use a recent deep learning technique, Normalizing Flows, to compute PV quantiles. It is compared to a typical neural architecture, referred to as Long Short-Term Memory.
We developed an integrated forecast-driven strategy modeled as a min-max-min robust optimization problem with recourse that is solved using a Benders decomposition procedure.
Two main cutting plane algorithms used to address the two-stage RO unit commitment problems are compared: the Benders-dual cutting plane and the column-and-constraint generation algorithms.
The convergence is checked by ensuring a gap below a threshold between the final objective and the corresponding deterministic objective value.
A risk-averse parameter assessment selects the optimal robust parameters and the optimal conservative quantile for the deterministic planner. Both the NF-based and LSTM-based planners outperformed the deterministic planner with nominal point PV forecasts. The NF model outperforms the LSTM in forecast value as the planner using the NF quantiles achieved higher profit than the planner with LSTM quantiles. Finally, a dynamic risk-averse parameter selection strategy is built by taking advantage of the PV quantile forecast distribution and provides further improvements. 
The CCG procedure converges ten times faster than the BD algorithm in this case study and achieves better results. However, it does not always converge to an optimal solution.

Overall, the RO approach for both the BD and CCG algorithms allows finding a trade-off between conservative and risk-seeking policies by selecting the optimal robust optimization parameters, leading to improved economic benefits compared to the baseline. Therefore, offering a probabilistic guarantee for the robust solution. However, the deterministic planner with the relevant PV quantile achieved interesting results. It emphasizes the interest to consider a well-calibrated deterministic approach. Indeed, it is easy to implement, computationally tractable for large-scale problems, and less prone to convergence issues. Note: this approach can be used in any other case study. It only requires a few months of data, renewable generation, and weather forecasts to train the forecasting models to compute reliable forecasts for the planner.

Several extensions are under investigation: (1) a stochastic formulation of the planner with improved PV scenarios based on Gaussian copula methodology or generated by a state-of-the-art deep learning technique such as Normalizing Flows, Generative Adversarial Networks or Variational AutoEncoders; (2) an improved dynamic risk-averse parameter selection strategy based on a machine learning tool capable of better-taking advantage of the PV quantiles distribution.

\section*{Acknowledgment}
The authors thank Quentin Louveaux, associate professor at Li\`ege University, for helping to propose demonstration \ref{demonstration}. In addition, the authors would like to thank the editor and the reviewers for the comments that helped improve the paper. 
Antoine Wehenkel, recipient of a F.R.S.- FNRS fellowship, and Xavier Fettweis, FNRS research associate, acknowledge the financial support of the FNRS (Belgium). Antonio Sutera is supported via the Energy Transition Funds project EPOC 2030-2050 organized by the FPS economy, S.M.E.s, Self-employed and Energy. 
\bibliographystyle{IEEEtran}
\bibliography{biblio}

\begin{thebibliography}{10}
\providecommand{\url}[1]{#1}
\csname url@samestyle\endcsname
\providecommand{\newblock}{\relax}
\providecommand{\bibinfo}[2]{#2}
\providecommand{\BIBentrySTDinterwordspacing}{\spaceskip=0pt\relax}
\providecommand{\BIBentryALTinterwordstretchfactor}{4}
\providecommand{\BIBentryALTinterwordspacing}{\spaceskip=\fontdimen2\font plus
\BIBentryALTinterwordstretchfactor\fontdimen3\font minus
  \fontdimen4\font\relax}
\providecommand{\BIBforeignlanguage}[2]{{%
\expandafter\ifx\csname l@#1\endcsname\relax
\typeout{** WARNING: IEEEtran.bst: No hyphenation pattern has been}%
\typeout{** loaded for the language `#1'. Using the pattern for}%
\typeout{** the default language instead.}%
\else
\language=\csname l@#1\endcsname
\fi
#2}}
\providecommand{\BIBdecl}{\relax}
\BIBdecl

\bibitem{birge2011introduction}
J.~R. Birge and F.~Louveaux, \emph{Introduction to stochastic
  programming}.\hskip 1em plus 0.5em minus 0.4em\relax Springer Science \&
  Business Media, 2011.

\bibitem{dumas2020probabilistic}
J.~Dumas, B.~Cornélusse, X.~Fettweis, A.~Giannitrapani, S.~Paoletti, and
  A.~Vicino, ``Probabilistic forecasting for sizing in the capacity firming
  framework,'' in \emph{2021 IEEE Madrid PowerTech}, 2021, pp. 1--6.

\bibitem{n2020controle}
\BIBentryALTinterwordspacing
A.~N'Goran, ``Contr{\^o}le optimal et gestion {\'e}nerg{\'e}tique d'une station
  d'{\'e}nergie autonome par optimisation robuste,'' Ph.D. dissertation, 2020,
  thèse de doctorat dirigée par Demassey, Sophie Contrôle, optimisation,
  prospective Université Paris sciences et lettres 2020. [Online]. Available:
  \url{http://www.theses.fr/2020UPSLM050}
\BIBentrySTDinterwordspacing

\bibitem{haessig2014dimensionnement}
\BIBentryALTinterwordspacing
P.~Haessig, ``Dimensionnement et gestion d’un stockage d’énergie pour
  l'atténuation des incertitudes de production éolienne,'' Ph.D.
  dissertation, 2014, thèse de doctorat dirigée par Multon, Bernard
  Électronique, électrotechnique, automatique Cachan, Ecole normale
  supérieure 2014. [Online]. Available:
  \url{http://www.theses.fr/2014DENS0030}
\BIBentrySTDinterwordspacing

\bibitem{parisio2016stochastic}
A.~Parisio, E.~Rikos, and L.~Glielmo, ``Stochastic model predictive control for
  economic/environmental operation management of microgrids: An experimental
  case study,'' \emph{Journal of Process Control}, vol.~43, pp. 24--37, 2016.

\bibitem{ben2009robust}
A.~Ben-Tal, L.~El~Ghaoui, and A.~Nemirovski, \emph{Robust optimization}.\hskip
  1em plus 0.5em minus 0.4em\relax Princeton University Press, 2009, vol.~28.

\bibitem{bertsimas2011theory}
D.~Bertsimas, D.~B. Brown, and C.~Caramanis, ``Theory and applications of
  robust optimization,'' \emph{SIAM review}, vol.~53, no.~3, pp. 464--501,
  2011.

\bibitem{bertsimas2012adaptive}
D.~Bertsimas, E.~Litvinov, X.~A. Sun, J.~Zhao, and T.~Zheng, ``Adaptive robust
  optimization for the security constrained unit commitment problem,''
  \emph{IEEE transactions on power systems}, vol.~28, no.~1, pp. 52--63, 2012.

\bibitem{jiang2011robust}
R.~Jiang, J.~Wang, and Y.~Guan, ``Robust unit commitment with wind power and
  pumped storage hydro,'' \emph{IEEE Transactions on Power Systems}, vol.~27,
  no.~2, pp. 800--810, 2011.

\bibitem{benders1962partitioning}
J.~F. Benders, ``Partitioning procedures for solving mixed-variables
  programming problems,'' \emph{Numerische mathematik}, vol.~4, no.~1, pp.
  238--252, 1962.

\bibitem{zhao2012robust}
L.~Zhao and B.~Zeng, ``Robust unit commitment problem with demand response and
  wind energy,'' in \emph{2012 IEEE power and energy society general
  meeting}.\hskip 1em plus 0.5em minus 0.4em\relax IEEE, 2012, pp. 1--8.

\bibitem{zeng2013solving}
B.~Zeng and L.~Zhao, ``Solving two-stage robust optimization problems using a
  column-and-constraint generation method,'' \emph{Operations Research
  Letters}, vol.~41, no.~5, pp. 457--461, 2013.

\bibitem{bottieau2019very}
J.~Bottieau, L.~Hubert, Z.~De~Gr{\`e}ve, F.~Vall{\'e}e, and J.-F. Toubeau,
  ``Very-short-term probabilistic forecasting for a risk-aware participation in
  the single price imbalance settlement,'' \emph{IEEE Transactions on Power
  Systems}, vol.~35, no.~2, pp. 1218--1230, 2019.

\bibitem{savelli2018new}
I.~Savelli, B.~Corn{\'e}lusse, A.~Giannitrapani, S.~Paoletti, and A.~Vicino,
  ``A new approach to electricity market clearing with uniform purchase price
  and curtailable block orders,'' \emph{Applied energy}, vol. 226, pp.
  618--630, 2018.

\bibitem{bertsimas2018scalable}
D.~Bertsimas and S.~Shtern, ``A scalable algorithm for two-stage adaptive
  linear optimization,'' 2018.

\bibitem{zhao2012exact}
L.~Zhao and B.~Zeng, ``An exact algorithm for two-stage robust optimization
  with mixed integer recourse problems,'' \emph{submitted, available on
  Optimization-Online. org}, 2012.

\bibitem{wang2015robust}
R.~Wang, P.~Wang, and G.~Xiao, ``A robust optimization approach for energy
  generation scheduling in microgrids,'' \emph{Energy Conversion and
  Management}, vol. 106, pp. 597--607, 2015.

\bibitem{rezende2015variational}
D.~Rezende and S.~Mohamed, ``Variational inference with normalizing flows,'' in
  \emph{International Conference on Machine Learning}.\hskip 1em plus 0.5em
  minus 0.4em\relax PMLR, 2015, pp. 1530--1538.

\bibitem{papamakarios2021normalizing}
G.~Papamakarios, E.~Nalisnick, D.~J. Rezende, S.~Mohamed, and
  B.~Lakshminarayanan, ``Normalizing flows for probabilistic modeling and
  inference,'' \emph{Journal of Machine Learning Research}, vol.~22, no.~57,
  pp. 1--64, 2021.

\bibitem{oord2018parallel}
A.~Oord, Y.~Li, I.~Babuschkin, K.~Simonyan, O.~Vinyals, K.~Kavukcuoglu,
  G.~Driessche, E.~Lockhart, L.~Cobo, F.~Stimberg \emph{et~al.}, ``Parallel
  wavenet: Fast high-fidelity speech synthesis,'' in \emph{International
  conference on machine learning}.\hskip 1em plus 0.5em minus 0.4em\relax PMLR,
  2018, pp. 3918--3926.

\bibitem{albergo2021introduction}
M.~S. Albergo, D.~Boyda, D.~C. Hackett, G.~Kanwar, K.~Cranmer, S.~Racanière,
  D.~J. Rezende, and P.~E. Shanahan, ``Introduction to normalizing flows for
  lattice field theory,'' 2021.

\bibitem{ge2020modeling}
L.~Ge, W.~Liao, S.~Wang, B.~Bak-Jensen, and J.~R. Pillai, ``Modeling daily load
  profiles of distribution network for scenario generation using flow-based
  generative network,'' \emph{IEEE Access}, vol.~8, pp. 77\,587--77\,597, 2020.

\bibitem{wehenkel2019unconstrained}
A.~Wehenkel and G.~Louppe, ``Unconstrained monotonic neural networks,'' in
  \emph{Advances in Neural Information Processing Systems}, 2019, pp.
  1545--1555.

\bibitem{fettweis2017reconstructions}
X.~Fettweis, J.~Box, C.~Agosta, C.~Amory, C.~Kittel, C.~Lang, D.~van As,
  H.~Machguth, and H.~Gall{\'e}e, ``Reconstructions of the 1900--2015 greenland
  ice sheet surface mass balance using the regional climate {MAR} model,''
  \emph{Cryosphere (The)}, vol.~11, pp. 1015--1033, 2017.

\bibitem{gneiting2007strictly}
T.~Gneiting and A.~E. Raftery, ``Strictly proper scoring rules, prediction, and
  estimation,'' \emph{Journal of the American statistical Association}, vol.
  102, no. 477, pp. 359--378, 2007.

\bibitem{zamo2018estimation}
M.~Zamo and P.~Naveau, ``Estimation of the continuous ranked probability score
  with limited information and applications to ensemble weather forecasts,''
  \emph{Mathematical Geosciences}, vol.~50, no.~2, pp. 209--234, 2018.

\bibitem{lauret2019verification}
P.~Lauret, M.~David, and P.~Pinson, ``Verification of solar irradiance
  probabilistic forecasts,'' \emph{Solar Energy}, vol. 194, pp. 254--271, 2019.

\bibitem{rahmaniani2017benders}
R.~Rahmaniani, T.~G. Crainic, M.~Gendreau, and W.~Rei, ``The benders
  decomposition algorithm: A literature review,'' \emph{European Journal of
  Operational Research}, vol. 259, no.~3, pp. 801--817, 2017.

\bibitem{lin2013exact}
H.~Lin and H.~{\"U}ster, ``Exact and heuristic algorithms for data-gathering
  cluster-based wireless sensor network design problem,'' \emph{IEEE/ACM
  transactions on networking}, vol.~22, no.~3, pp. 903--916, 2013.

\end{thebibliography}

\section{Appendix: Forecasting techniques}\label{appendix:forecasting}

The processes described in Section \ref{sec:capacity_firming_process} require day-ahead and intraday \textit{top-quality} forecasts. The more accurate the forecasts, the better the planning and the control. The robust optimization-based approach needs quantile forecasts to define the uncertainty interval. 
To this end, the Normalizing Flows technique is used to compute quantile day-ahead forecasts compared to a common alternative technique using a Long Short-Term Memory neural network. The controller requires intraday point forecasts that are computed by an encoder-decoder architecture.
Appendices \ref{sec:nfs} and \ref{sec:lstm} introduce the NF and LSTM techniques implemented.
Appendix \ref{sec:forecast_quality} proposes a quality evaluation of the NF and LSTM PV quantiles.

\subsection{Normalizing Flows}\label{sec:nfs}

We investigate the use of \textit{Normalizing Flows} \cite{rezende2015variational} that are a promising method for modeling stochastic generative processes. NFs define a new class of probabilistic generative models. It has gained increasing interest from the deep learning community. 
They have proven to be an effective way to model complex data distributions with neural networks in many domains such as image, video, and audio generation \cite{papamakarios2021normalizing}, speech synthesis \cite{oord2018parallel}, or fundamental physics \cite{albergo2021introduction}.

\begin{figure}[tb]
	\centering
	\includegraphics[width=\linewidth]{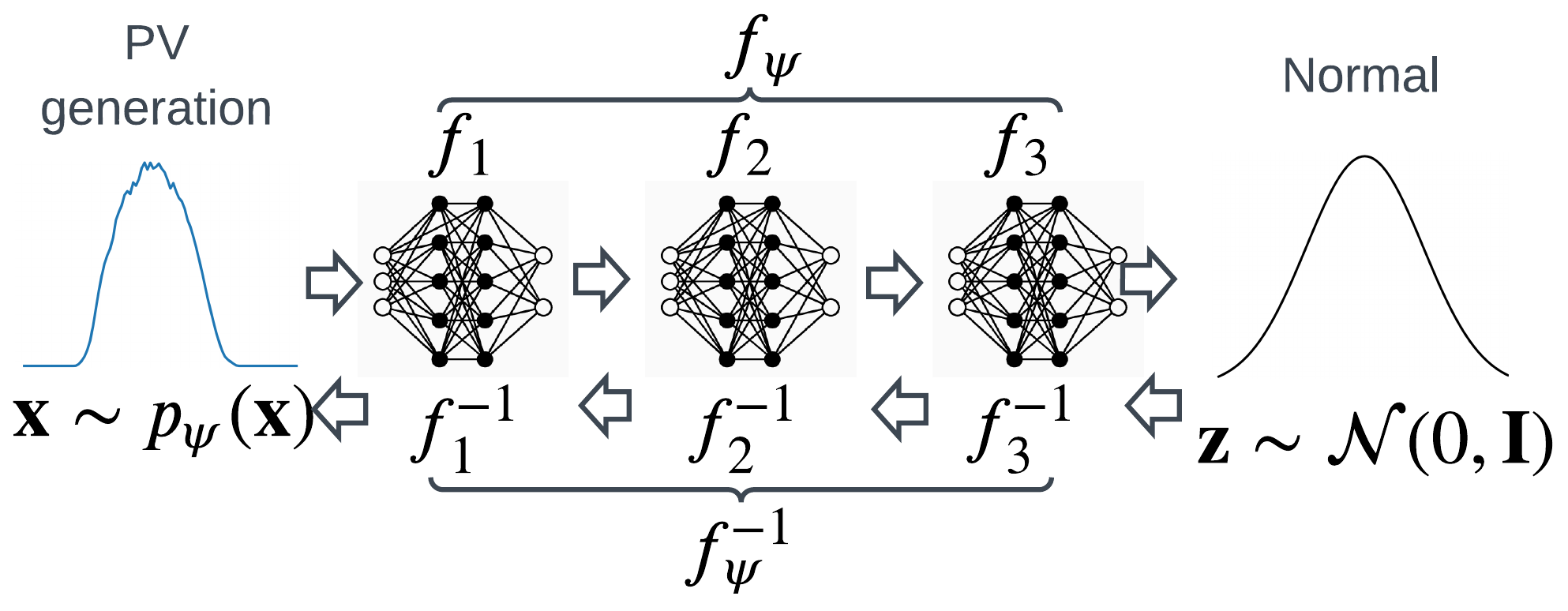}
	\caption{A three-step NF.}
	\label{fig:nf_process}
\end{figure}

In this Appendix, let $x$ be the random variable of interest, \textit{i.e.}, the PV generation.
Normalizing Flows, such as depicted in Figure \ref{fig:nf_process}, are defined as a sequence of invertible transformations $f_k : \mathbb{R}^T  \rightarrow \mathbb{R}^T$, $k = 1, \ldots, K$, composed together to create an expressive invertible mapping $f_\psi := f_1 \circ \ldots  \circ f_K : \mathbb{R}^T  \rightarrow \mathbb{R}^T$. This composed function can be used to perform density estimation, using $f_\psi$ to map a sample $\mathbf{x} \in  \mathbb{R}^T $ onto a latent vector $\mathbf{z} \in  \mathbb{R}^T $ equipped with a known and tractable probability density function $p_z$, \textit{e.g.}, a \textit{Normal} distribution. The transformation $f_\psi$ implicitly defines a density $p_\psi(\mathbf{x})$ that is given by the change of variables
\begin{align}
\label{eq:change_formula}	
p_\psi(\mathbf{x})  & = p_z(f_\psi(\mathbf{x}))|\det J_{f_\psi}(\mathbf{x})| , 
\end{align}
where $ J_{f_\psi}$ is the Jacobian of $f_\psi$ regarding $\mathbf{x}$. The model is trained by maximizing the log-likelihood $\sum_{i=1}^N \log p_\psi(\mathbf{x}^i)$ of the model's parameters $\psi$ given the dataset $\mathcal{D}$. 
The motivations to use NFs over more traditional deep learning approaches are three-fold:
\begin{enumerate}
    \item Evaluating NFs in power system applications in terms of forecast value, and more particularly in the capacity firming framework. To the best of our knowledge, only \cite{ge2020modeling} used NFs to generate daily load profiles. However, the model did not consider the weather forecasts, and the assessment is only performed on the forecast quality. In contrast, we implement a conditional NF to compute improved weather-based PV forecasts.
    \item NFs can challenge state-of-the-art deep learning techniques in terms of quality, as demonstrated by \cite{ge2020modeling}. Our study demonstrates that they are also more accurate in forecast value and can be used effectively by non-expert deep learning practitioners.
	\item NFs can directly be trained by maximum likelihood estimation. Therefore, in contrast to other deep learning generative models, \textit{e.g.}, Generative Adversarial Networks (GANs) or Variational AutoEncoders (VAEs), NFs provide access to the exact likelihood of the model’s parameters.
\end{enumerate}
There are many possible implementations of NFs, see \cite{papamakarios2021normalizing} for a comprehensive review.
In this paper, the class of Affine Autoregressive flows is implemented\footnote{\url{https://github.com/AWehenkel/Normalizing-Flows} \cite{wehenkel2019unconstrained}}. A five-step Affine Autoregressive flow is trained by maximum likelihood estimation with 500 epochs and a learning rate set to $10^{-4}$.

\subsection{Long Short Term Memory}\label{sec:lstm}

The NF probabilistic day-ahead forecasts are compared to one of the most famous deep learning techniques adopted in energy forecasting applications: a Long Short-Term Memory neural network. More particularly the neural network implemented is composed of a LSTM cell and feed-forward layers, and is referred to as LSTM in the rest of the paper. The number of LSTM units is $n_\text{input} + (n_\text{output} - n_\text{input}) / 3$, and the number of neurons of the feed-forward layer $n_\text{input} + 2 \times (n_\text{output} - n_\text{input}) / 3$, with $n_\text{input}$ and $n_\text{output}$ the number of neurons of the input and output layers, respectively. The activation functions are the ReLU, the learning rate is set to $10^{-3}$, the number of epoch to 500 with a batch size of 64. The model is trained by quantile regression that consists of minimizing the quantile loss over the dataset.

\subsection{Encoder-Decoder}\label{sec:ed}

The intraday point forecasts are computed by an innovative architecture, referred to as encoder-decoder \cite{bottieau2019very}. It comprises two different networks and has recently shown promising results for translation tasks, speech recognition applications, and imbalance price forecasting. The encoder-decoder processes features from the past, such as past PV observations, to extract the relevant historical information contained into a reduced vector of fixed dimensions, based on the last hidden state. Then, the decoder processes this representation along with the known future information such as weather forecasts.
This architecture is implemented with a LSTM as the encoder and a multilayer perceptron as the decoder. The encoder has $2 \times n_\text{input}$ units with $n_\text{input}$ the number of neurons of the encoder input layer, features from the past. Then, the encoder output is merged with the weather forecasts becoming the decoder input layer that has $n_\text{output} /2$ neurons. The activation functions are the ReLU. The learning rate is $10^{-3}$, and the number of epoch to 500 with a batch size of 64. The model is trained by minimizing the mean squared error over the dataset.

\subsection{Quantile forecasts quality evaluation}\label{sec:forecast_quality}

The \textit{quantile score}, \textit{reliability diagram}, and \textit{continuous ranked probabilistic score} are used to assess the quantile forecast quality of both the NF and LSTM models. Forecast quality corresponds to the ability of the forecasts to genuinely inform of future events by mimicking the characteristics of the processes involved. Forecast value relates, instead, to the benefits from using forecasts in a decision-making process, such as participation in the electricity market. In this Appendix, we focus only on the forecast quality evaluation.

Both NF and LSTM models use the weather forecasts of the MAR (Regional Atmosphere Model) regional climate model provided by the Laboratory of Climatology of the Li\`ege University \cite{fettweis2017reconstructions}. The NF model generates day-ahead scenarios, and the quantiles are derived. The LSTM model computes the quantiles directly as it is trained by minimizing the quantile loss. The set of PV quantiles considered for the assessment is $\mathcal{Q} =  \{q=10 \%, \ldots, 90 \%\}$.

The continuous ranked probability score (CRPS) \cite{gneiting2007strictly} penalizes the lack of resolution of the predictive distributions as well as biased forecasts. It is negatively oriented, \textit{i.e.}, the lower, the better, and for point forecasts, it turns out to be the mean absolute error.
Let $\hat{p}^q_{t+k|t}$ be the PV quantile forecast $q$ generated at time $t$ for lead time $t+k$. The energy form of the CRPS for lead time $k$ is estimated following \cite{zamo2018estimation} over the dataset $\mathcal{D}$ of length $N$ $\forall k=k_1, \ldots, k_T$ as follows
\begin{align}\label{eq:CRPS_eNRG}
\text{CRPS}(k) & = \frac{1}{N}\sum_{t \in \mathcal{D}} \big [ \frac{1}{Q}\sum_{q = 1}^Q|\hat{p}^q_{t+k|t}-p_{t+k}| \notag \\
& - \frac{1}{2 Q^2} \sum_{q,q' = 1}^Q |\hat{p}^q_{t+k|t} - \hat{p}^{q'}_{t+k|t}|  \big].
\end{align}
The quantile score (QS) is complementary to the CRPS as it permits obtaining detailed information about the forecast quality at specific probability levels, \textit{i.e.}, over-forecasting or under-forecasting, and particularly those related to the tails of the predictive distribution \cite{lauret2019verification}. It is negatively oriented and assigns asymmetric weights to negative and positive errors for each quantile. The quantile score for quantile $q$ is estimated over the dataset $\mathcal{D}$ of length $N$ and for all lead times $k$ as follows
\begin{subequations}
\begin{align}\label{eq:quantile_score}
QS(q) = & \frac{1}{N}\sum_{t \in \mathcal{D}} \frac{1}{T}\sum_{k=k_1}^{k_T} \rho_q(\hat{p}^q_{t+k|t}, p_{t+k})  ,\\
\rho_q(\hat{p}, p) := & \max \big\{(1-q) (\hat{p} - p), q (p - \hat{p}) \big\}.
\end{align}
\end{subequations}
Finally, the reliability diagram is a visual verification used to evaluate the reliability of the quantiles derived from the scenarios. Quantile forecasts are reliable if their nominal proportions are equal to the proportions of the observed value.
\begin{figure}[htbp]
	\centering
	\begin{subfigure}{.25\textwidth}
		\centering
		\includegraphics[width=\linewidth]{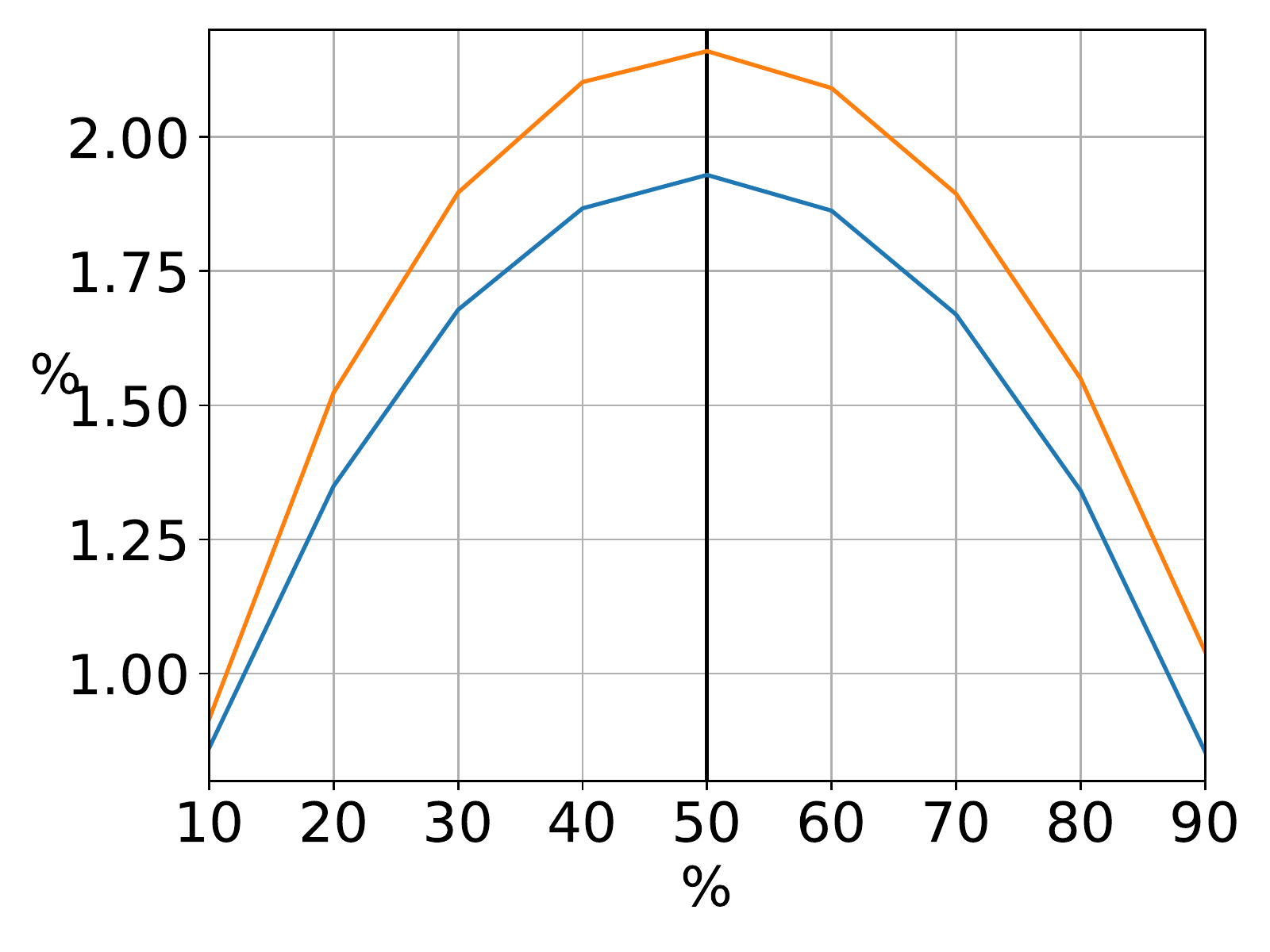}
		\caption{$QS(q)$.}
	\end{subfigure}%
	\begin{subfigure}{.25\textwidth}
		\centering
		\includegraphics[width=\linewidth]{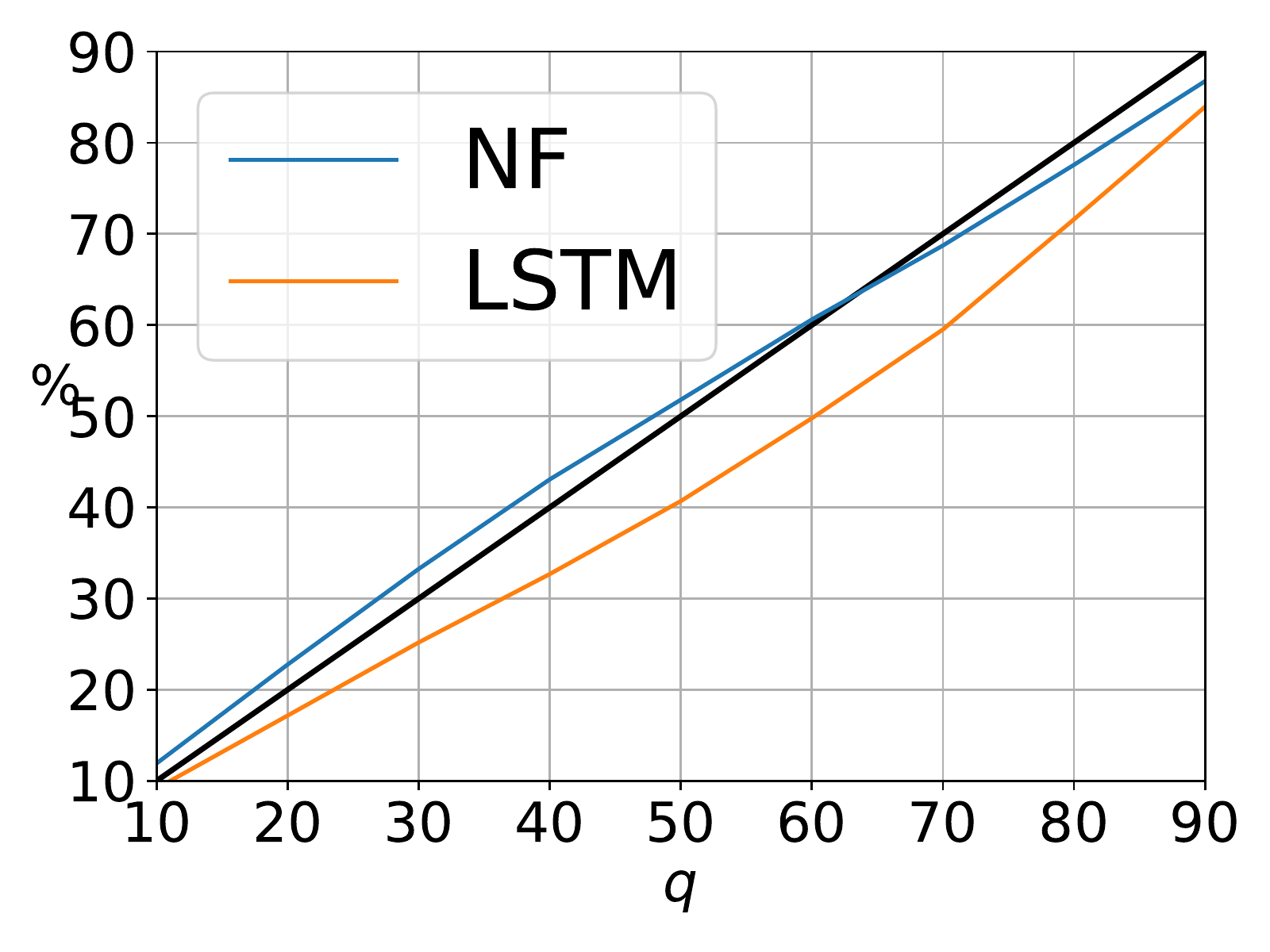}
		\caption{Reliability.}
	\end{subfigure}
	\begin{subfigure}{.25\textwidth}
		\centering
		\includegraphics[width=\linewidth]{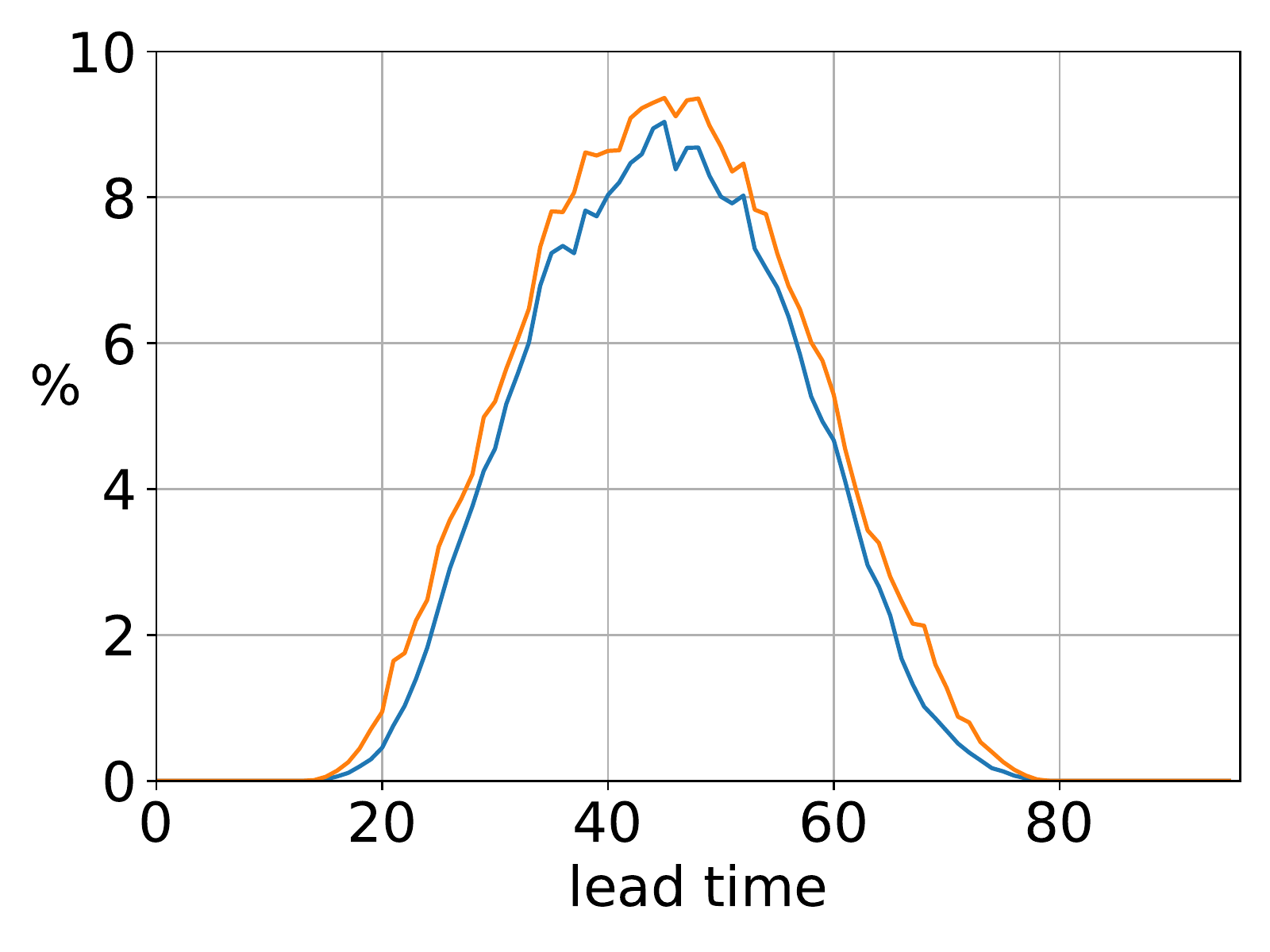}
		\caption{$CRPS(k)$.}
	\end{subfigure}%
	\caption{Quantile forecast quality evaluation.}
	\label{fig:forecast_eval}
\end{figure}

Figure \ref{fig:forecast_eval} provides the results for these quality metrics computed over the entire dataset normalized by the total installed capacity. The NF model outperforms the LSTM model with average values of 1.49 \% and 2.80 \% \textit{vs.} 1.69 \% and 3.15 \% for the QS and CRPS, respectively. The NF quantiles are also more reliable, as indicated by the reliability diagram. These results motivate the use of the NFs as they outperform common deep learning approaches such as LSTM models. However, the core focus of this paper is robust optimization in the capacity framework. Therefore, an extensive NF assessment in forecast quality and value compared with state-of-the-art deep learning models is out of the scope and will be proposed in another study.

\section{Appendix: Benders dual cutting plane warm-start procedure}\label{appendix:warm-start}

Section \ref{sec:warm_start} presents the BD warm-start procedure implemented and Section \ref{sec:res-BD-warm-start} the improvement in terms of computation time provided by the initial set of cuts. 

\subsection{BD warm-start procedure}\label{sec:warm_start}

A four-dimension taxonomy of algorithmic enhancements and acceleration strategies is proposed by \cite{rahmaniani2017benders}: solution generation, solution procedure, decomposition strategy, cut generation. The solution generation is the method used to set trial values for the SP. The quality of these solutions impacts the number of iterations, as the SP uses them to generate cuts and bounds. The standard strategy is to solve the MP without modification. However, heuristics can be used as a warm-start strategy to generate an initial set of tight cuts to strengthen the MP. A simple heuristic is proposed by \cite{lin2013exact} to generate feasible solutions and a set of good initial cuts. Computational evidence demonstrated the efficiency of this approach in terms of solution quality and time. 
Therefore, we designed the following warm-start method to improve the Benders convergence by building an initial set of cuts $\{\theta_i\}_{1\leq i \leq I}$ for the master problem (\ref{eq:master_problem}). It consists of sampling renewable generation trajectories that are assumed to be close to the worst trajectory of $\mathcal{P}$. 
%
Let $t_1$ and $t_f$ be the time periods corresponding to the first and last non null 50 \% quantile values. If $m = t_f -(t_1 + \Gamma-1) > 0 $, $m$ trajectories are sampled. The $m^\text{th}$ sampled trajectory is built by setting the $\Gamma$ values of the 50 \% quantile to the $p_t^{min}$ lower bound for time periods $t_1 + (m-1)\leq t \leq t_1 + \Gamma -1 + (m-1)$.
An additional trajectory is built by setting the $\Gamma$ maximum values of the 50  \% quantile to $p_t^{min}$ lower bound. 
%
Then, for each sampled trajectory $p_{t,i}$, the MILP formulation (\ref{eq:det_compact}) is used to compute the related engagement plan $x_{t,i}$. Finally, the cut $\theta_i$ is built by solving (\ref{eq:worst_case_dispatch}) where the uncertainty set is a singleton $\mathcal{P} =\{ p_{t,i }\}$, and the engagement plan is $x_{t,i}$ to retrieve the optimal values with (\ref{eq:optimality_cut}): $\theta_i = G \big(x_{t,i}, \alpha_{t,i}, \phi_{t,i} \big),  \ i = 1 \ldots I$.

\subsection{BD convergence warm-start improvement}\label{sec:res-BD-warm-start}

\begin{figure}[tb]
		\centering
	\begin{subfigure}{.25\textwidth}
		\centering
	\includegraphics[width=\linewidth]{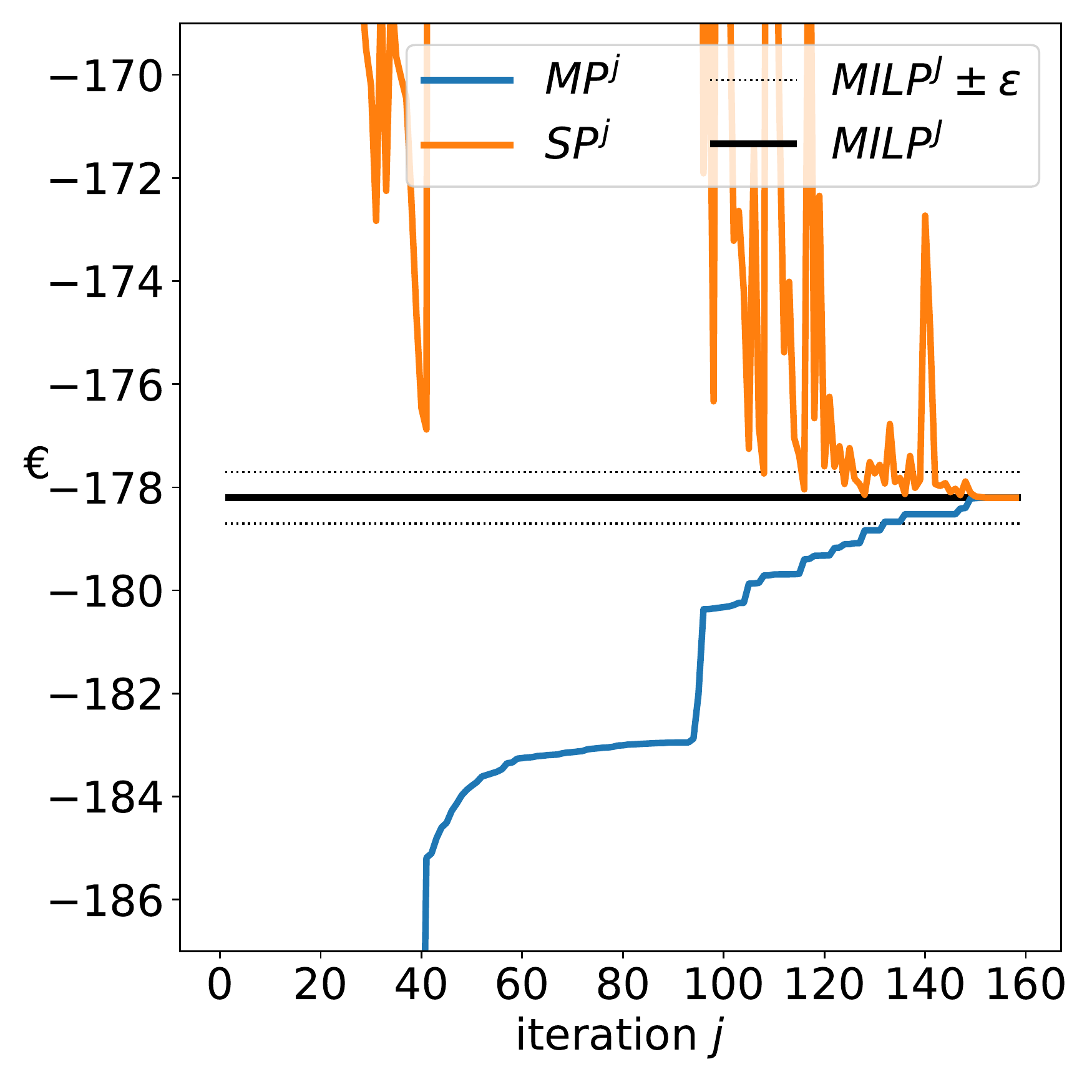}
	\end{subfigure}%
	\begin{subfigure}{.25\textwidth}
		\centering
	\includegraphics[width=\linewidth]{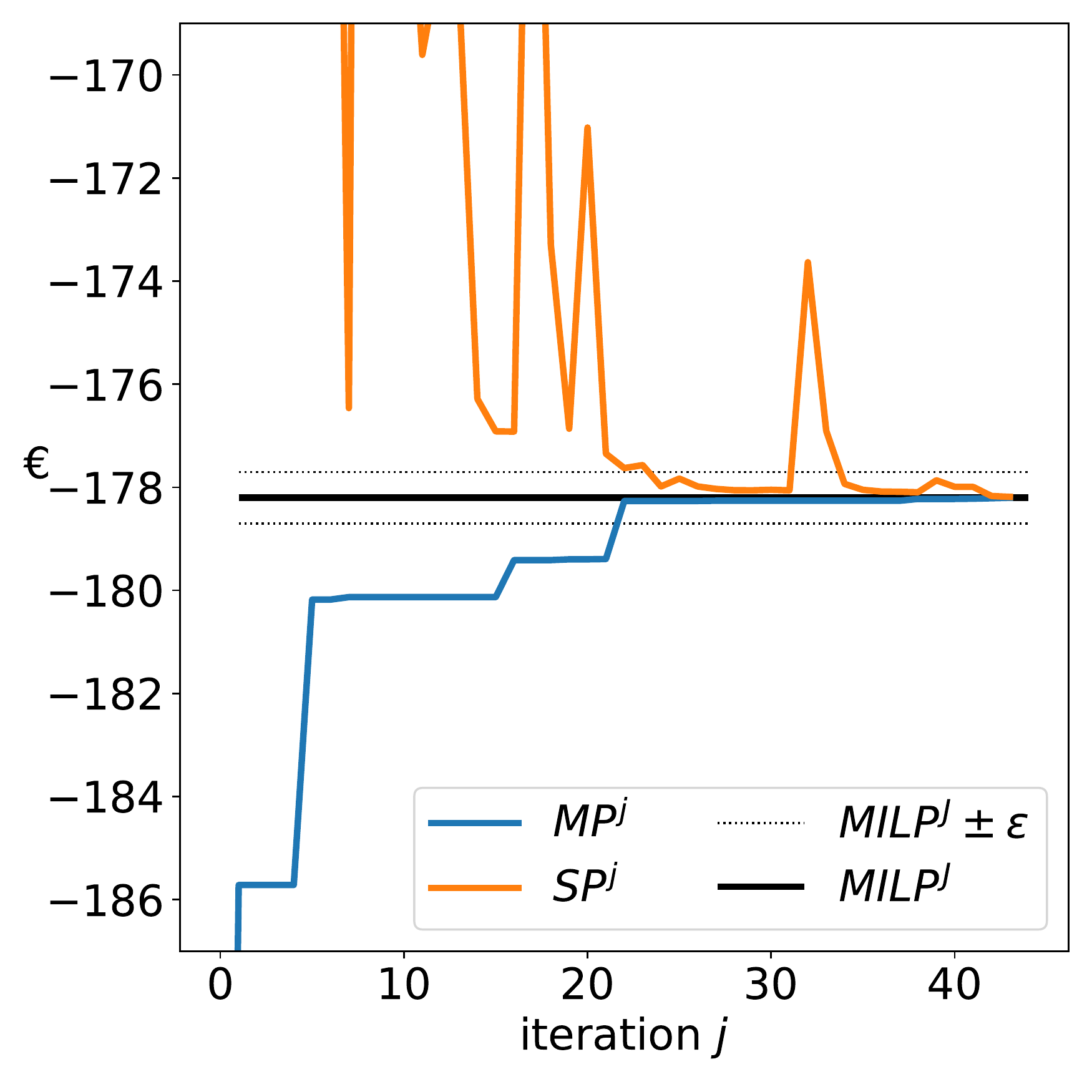}
	\end{subfigure}
	\caption{BD convergence without (left) and with (right) warm-start on $\convdate$.}
	\label{fig:conv_res}
\end{figure}

Overall, the warm-start procedure of the BD algorithm improves the convergence by reducing the number of iterations and allows to reach more frequently optimal solutions. In addition, it reduces the number of times the big-M's values need to be increased before reaching the final convergence criterion with the MILP.
It is illustrated by considering the dynamic risk-averse parameters strategy with $[d_\Gamma, d_q] = [10,10]$.
Figure \ref{fig:conv_res} depicts the reduction of the total number of iterations $J$ required to converge below the threshold $\epsilon$, on a specific day of the dataset. It is divided by 3.6 from 159 to 44. The computation time is divided by 4.1 from 7.4 min to 1.8 min. 
Table \ref{tab:computation_times} provides the computation times (min) statistics over the entire dataset with and without warm-start. The averaged $t^{av}$ and total $t^{tot}$ computation times are drastically reduced when using the warm-start.
\begin{table}[tb]
\renewcommand{\arraystretch}{1.25}
	\begin{center}
		\begin{tabular}{lrrrrr}
			\hline \hline
			Warm-start & $t^{av}$ & $t^{50\%}$ & $t^{min}$ & $t^{max}$ & $t^{tot}$ \\ \hline
			False & 3.5 & 2.0 & $<0.1$ & 34.1  & 105.4 \\
			True  & 2.0 & 0.7 & $<0.1$ & 30.4  & 61.3  \\ \hline \hline
		\end{tabular}
		\caption{Computation times (min) statistics.}
		\label{tab:computation_times}
	\end{center}
\end{table}

\section{Appendix: column and constraints generation algorithm}\label{appendix:ccg}

We implemented the column and constraints generation procedure proposed by \cite{zhao2012robust,zeng2013solving}.
The following master problem ($\text{MP}_2$) is solved at iteration $j$
\begin{subequations}\label{eq:ccg-MP}	
	\begin{align}
	\min_{x_t \in \mathcal{X}, \ \theta, \ \{y_t^s\}_{0 \leq s\leq j}} &    \ \theta \\
	& \theta \geq J \big(x_t, y_t^s \big),  \quad s = 0 \ldots j \label{eq:ccg-optimality-cut} \\
	&  y_t^s\in \Omega(x_t, \hat{p}_t^{\star,s}), \quad s = 0 \ldots j , \label{eq:ccg-feasability-cut}
	\end{align}
\end{subequations}
where constraints (\ref{eq:ccg-optimality-cut}) and (\ref{eq:ccg-feasability-cut}) serve as optimality and feasibility, respectively. $\{y_t^s\}_{0 \leq s\leq j}$ are the new variables added to the $\text{MP}_2$, and $\hat{p}_t^{\star,s}$ represent the worst PV trajectory computed by the SP at iteration $0 \leq s\leq j$.
Note: in our CCG implementation, we solve the SP with the same approach as the SP of the BD algorithm.
Figure \ref{algo-ccg} depicts the CCG algorithm implemented that is similar to the BD procedure. 
The initialization step consists of setting the initial big-M's values $M_t^- = 1$ and $M_t^+ = 0 \ \forallt$, the time limit resolution of the sub-problem (\ref{eq:worst_case_dispatch}) to 10 s, and the threshold convergence $\epsilon $ to 0.5 \euro. Let $MP_2^j $, $SP^j$, be the $\text{MP}_2$ and SP objective values at iteration $j$, the lower and upper bounds, respectively, and $MILP^J$ the MILP objective value using the worst renewable generation trajectory $\hat{p}_t^{\star,J}$ at the last CCG algorithm iteration $J$.
Note: there is a maximum of 50 iterations between the SP and the $\text{MP}_2$ before checking the convergence with the MILP. If the criterion is not reached, the big-M's values are increased. Indeed, at each iteration $j$ the $y_t^j$ variables are added to the $\text{MP}_2$. In our case, it represents approximately 1 000 new variables at each iteration. With 50 iterations, the $\text{MP}_2$ is a MILP with approximately 50 000 variables which begins to be hard to solve within a reasonable amount of time.
\begin{figure}
	\begin{algorithmic}
		\STATE Initialization.
		\WHILE{$|MILP^J - MP_2^J |>  \epsilon$ and $M_t^- <500$ }
		\STATE Initialize $j=0$, solve the $MP_2$ (\ref{eq:ccg-MP}) and retrieve $x_{t,0}$.
		\WHILE{the two last $|MP_2^j - SP^j| $ are not $< \epsilon$ and $j < 50$ }
		\STATE Solve the SP (\ref{eq:worst_case_dispatch}) with $x_{t,j}$ as parameters:
		\STATE Create variables $y_t^j$ in $MP_2$.
		\STATE Retrieve $\hat{p}_t^{\star,j}$ from the SP, and add the feasibility cut to the $MP_2$: $y_t^j\in \Omega(x_t, \hat{p}_t^{\star,j})$.
		\IF{the SP is bounded}
		\STATE Add the optimality cut: $ \theta \geq J \big(x_t, y_t^j \big)$.
		\STATE Update the upper bound: $SP^j= R(x_{t,j})$.
		\STATE SP check: no simultaneous charge and discharge.
		\ENDIF
		\STATE Solve the $MP_2$ (\ref{eq:ccg-MP}): get the optimal values $\theta_j, x_{t,j}$.
		\STATE Update the lower bound: $MP^j_2=\theta_j$ and $j = j+1$.
		\ENDWHILE
		\STATE $j=J$: convergence between the SP and MP is reached. Check convergence with MILP: get $\hat{p}_t^{\star,J}$ from $SP^J$ and compute $MILP^J$ (\ref{eq:det_compact}).
		\IF{$|MILP^J - MP^J_2 |>  \epsilon$}
		\IF{$M_t^- \leq 50$}
		\STATE Update big-M's values $M_t^- = 10 + M_t^- \ \forallt$. 
		\ELSE
		\STATE Update big-M's values $M_t^- = 100 + M_t^- \ \forallt$.
		\ENDIF
		\STATE Reset $j$ to 0 and restart algorithm with a new $MP_2$.
		\ENDIF
		\ENDWHILE
	\STATE Retrieve the final $x_{t,J}$ engagement.
	\end{algorithmic}
	\caption{Column and constraints generation algorithm.}\label{algo-ccg}
\end{figure}

\end{document}